\documentclass[%
  reprint,
 superscriptaddress,
  amsmath,amssymb,
  aps,
 ]{revtex4-2}
\usepackage[dvipdfmx]{graphicx}
\usepackage{epsfig}
\usepackage{times}
\usepackage{amsmath}
\usepackage{amsfonts}
\usepackage{amssymb}
\usepackage[usenames]{color}
\usepackage{graphicx}
\usepackage{comment}
\usepackage{bm}% bold math
\usepackage{xcolor}
\usepackage{braket}
\usepackage{here}
\usepackage{ulem}

 \newcommand{\ym}[1]{\textcolor{black}{#1}}
 
\newcommand{\hk}[1]
{\textcolor{black}{#1}}
\newcommand{\as}[1]{\textcolor{black}{#1}}

\begin{document}
%\preprint{APS/123-QED}

\title{
Characterization and
generation of a %metrologically advantageous
SQL-beating
catlike state %at finite temperature 
through repetitive measurements}
% Force line breaks with \\
%\thanks{A footnote to the article title}%

\author{Mamiko Tatsuta}
% \altaffiliation[Also at ]{Physics Department, XYZ University.}%Lines break automatically or can be forced with \\
\email[email address:]{mamikotatsuta@gmail.com}
\affiliation{
Department of Electrical, Electronic, and Communication Engineering,
Faculty of Science and Engineering, Chuo university,
1-13-27, Kasuga, Bunkyo-ku, Tokyo 112-8551, Japan
%National Institute of Advanced Industrial Science and Technology (AIST), 1-1-1 Umezono, Tsukuba, Ibaraki 305-8568, Japan
}
\author{Yuichiro Matsuzaki}
\email[email address:] {
ymatsuzaki872@g.chuo-u.ac.jp
}
\affiliation{
Department of Electrical, Electronic, and Communication Engineering,
Faculty of Science and Engineering, Chuo university,
1-13-27, Kasuga, Bunkyo-ku, Tokyo 112-8551, Japan
}
\author{Hiroki Kuji}%
\email[email address:] {1225702@ed.tus.ac.jp}
\affiliation{
Department of Electrical, Electronic, and Communication Engineering,
Faculty of Science and Engineering, Chuo university,
1-13-27, Kasuga, Bunkyo-ku, Tokyo 112-8551, Japan
}
\affiliation{
Department of Physics, Tokyo University of Science,
1-3 Kagurazaka, Shinjuku, Tokyo, 162-8601,  Japan
}

\author{Ryusuke Hamazaki}
\email[email address:]{ryusuke.hamazaki@riken.jp}
\affiliation{
Nonequilibrium Quantum Statistical Mechanics RIKEN Hakubi Research Team, RIKEN Pioneering Research Institute (PRI), Wako, Saitama 351-0198, Japan
}
\affiliation{
RIKEN Center for Interdisciplinary Theoretical and
Mathematical Sciences (iTHEMS), RIKEN, Wako 351-0198, Japan
}
\author{Akira Shimizu}%
\email[email address:]{shmz@as.c.u-tokyo.ac.jp}
\affiliation{Institute for Photon Science and Technology, The University of Tokyo, 7-3-1 Hongo, Bunkyo-ku, Tokyo 113-0033, Japan}
\affiliation{Center for Quantum Information and Quantum Biology, The University of Osaka, Toyonaka, Osaka 560-0043, Japan}

\date{\today}

\begin{abstract}
Sensitivity in metrology without entanglement is limited by the standard quantum limit (SQL).
Recent studies have found that the Heisenberg-limited scaling, the ultimate sensitivity in quantum metrology, can be achieved by generalized cat states, which are characterized by an index that indicates coherence among macroscopically distinct states and are associated with additive observables.
Although generalized cat states include diverse states, encompassing classical mixtures of exponentially large numbers of states,
the preparation of large generalized cat states has not been demonstrated yet.
Here we characterize SQL-beating catlike states using the index $q$ indicating macroscopic coherence and prove that any state with $q>1.5$ has a potential to surpass the SQL when used as a sensor. We propose a protocol to generate them through repetitive measurements on a quantum spin system of $N$ spins, which we call a spin ensemble. 
Starting from a thermal equilibrium state of the spin ensemble, we demonstrate that we can increase the coherence among the spin ensemble via repetitive weak measurements of its total magnetization, which is indirectly measured through an ancillary qubit collectively coupled to the ensemble. 
Notably, our method for creating the SQL-beating catlike states requires no dynamical control over the spin ensemble. 
As a potential experimental realization, we discuss a hybrid system composed of a superconducting flux qubit and donor spins in silicon. 
Our results pave the way for the realization of entanglement-enhanced quantum metrology in state-of-the-art technology.
\end{abstract}

\pacs{Valid PACS appear here} 
\maketitle

\section{Introduction}

\ym{High-precision measurement 
\hk{is of paramount} importance in both 
\hk{fundamental research} and applied 
\hk{science}~\cite{giovannetti2004quantum, giovannetti2011advances, taylor2016quantum, degen2017quantum}.
\hk{Numerous critical parameters}, including temperature, electric field, and pressure,
\hk{require precise measurement. In particular}, magnetic field sensing has garnered considerable attention~\cite{wineland1992spin, wineland1994squeezed, toth2014quantum}, serving not only to investigate material properties
but also to 
\hk{elucidate} biological 
\hk{mechanisms}. 
Substantial efforts have been devoted to %\ymdel{improving}
\hk{enhancing} the sensitivity of magnetic-field sensors~\cite{paris2009quantum, chin2012quantum, PhysRevLett.111.120401, jones2009magnetic, PhysRevLett.79.3865, kuzmich1998atomic, fleischhauer2000quantum, PhysRevLett.91.250801, leibfried2004toward, PhysRevLett.93.173002, Dunningham2006-zb, PhysRevA.84.012103, demkowicz2012elusive, bohnet2014reduced, PhysRevLett.115.170801, dooley2016hybrid, PhysRevLett.116.053601, PhysRevLett.120.140501}. 
One 
\hk{promising approach} to 
\hk{improving} sensitivity is 
\hk{the utilization of} qubits~\cite{PhysRevLett.31.273, PhysRevLett.89.130801, dang2010ultrahigh, bal2012ultrasensitive, Toida2019-eo, PhysRevB.80.115202, balasubramanian2009ultralong, dolde2011electric, ishikawa2012optical}. 
Magnetic fields 
\hk{induce shifts in} the resonant frequency of %\ymdel{the qubit}
\hk{qubits}, 
allowing  the 
\hk{estimation} of 
magnetic-field 
\hk{strength via} Ramsey-type measurements. 
\hk{When employing} $N$ qubits 
\hk{in} separable states\hk{,} the uncertainty (i.e., the reciprocal of sensitivity) scales as $N^{-1/2}$, \ym{which is called}
the standard quantum limit (SQL). 
\hk{Conversely, using} a 
\hk{specific} type of entangled state, the sensitivity scales as $N^{-1}$, 
\hk{achieving what} is 
\hk{known as} Heisenberg-limited scaling. 
In realistic 
\hk{scenarios}, 
\hk{environmental decoherence} cannot 
\hk{be avoided~\cite{chin2012quantum, jones2009magnetic, PhysRevA.84.012103, PhysRevLett.115.170801, dooley2016hybrid, PhysRevLett.120.140501,palma1996quantum, PhysRevLett.116.120801, PhysRevA.92.010102, zurek1991decoherence, Breuer2002-jn,tratzmiller2020limited}, making it challenging to reach} 
\hk{Heisenberg-limited scaling. Nevertheless}, it is 
\hk{established that under certain conditions of decoherence}, it is 
\hk{possible to surpass} the SQL in scaling. 
For 
\hk{instance}, 
\hk{in} the 
\hk{presence} of time-inhomogeneous dephasing, the uncertainty scales as $N^{-3/4}$, 
\hk{referred to as} the Zeno limit~\cite{matsuzaki2011magnetic,chin2012quantum}.
}

\ym{The concept of 
superposition 
\hk{involving} macroscopically distinct states has been \hk{a topic} of fundamental interest since its inception by 
\hk{Schr\"{o}dinger}~\cite{Schrodinger1935-oo}. 
\hk{The} Greenberger-Horne-Zeilinger (GHZ)~\cite{greenberger1990bell, PhysRevLett.106.130506, dicarlo2010preparation} state is 
\hk{a quintessential example of such a superposition}. The GHZ state is expressed as $(\ket{\uparrow}^{\otimes N}+\ket{\downarrow}^{\otimes N})/\sqrt{2}$, 
where $\ket{\uparrow}$ and $\ket{\downarrow}$ are eigenstates of the Pauli operator \textcolor{black}{$\hat{\sigma}_3$} with eigenvalues $+1$ and $-1$, respectively.
Although a great deal of effort has been devoted to producing a superposition of macroscopically distinct states \cite{monroe1996schrodinger,brune1996observing,friedman2000quantum,leibfried2005creation,ourjoumtsev2007generation,deleglise2008reconstruction,gao2010experimental,vlastakis2013deterministically,kirchmair2013observation,wang2016schrodinger,stojanovic2022interconversion,stojanovic2023dicke}, a unified criterion to 
\hk{determine} whether a given state contains such macroscopic superpositions remains elusive~\cite{RevModPhys.90.025004}. }

Among 
\hk{the various} potential metrics, an index $q$~\cite{PhysRevLett.95.090401}, a real number satisfying $1\leq q \leq 2$, is 
\hk{particularly noteworthy for gaining} deeper insights into the correlation between cat states and sensor technologies.
A state with $q=2$ is 
\hk{referred to as a generalized cat state}.
Notably, there exists a generalized cat state $\hat\rho_m$ with 
exponentially small purity, i.e., $\mathrm{Tr}(\hat\rho_m^2)=\exp(-\Theta(N))$~\footnote{We use the notation $\Theta(N^k)$
in addition to the standard notation $O(N^k)$, i.e., $O(N^k)/N^k \rightarrow \text{const.} (\neq 0)$  in the limit $N\rightarrow\infty$, as the following meaning:
For a function $g$ of $N$, we say $g = \Theta(N^k)$ if
$g/N^k$  approaches a positive constant as $N\rightarrow \infty$. If
$g = \Theta(N^k)$ then $g=O(N^k)$, but the inverse is not
necessarily true.}.
It 
\hk{has been shown that utilizing} generalized cat states 
\hk{for} magnetic-field
\hk{sensing} can achieve Heisenberg-limited scaling 
\hk{in the absence of decoherence~\cite{huelga1997improvement} and the Zeno limit in the presence of time-inhomogeneous dephasing~\cite{matsuzaki2011magnetic,chin2012quantum}}.
\hk{Thus,} generalized cat states 
\hk{present a compelling} metrological approach, leveraging quantum properties to \hk{significantly} enhance sensitivity.

\ym{
\hk{A} theoretical proposal
\hk{exists for the creation of a} generalized \hk{cat} state~\cite{tatsuta2018conversion}. 
\hk{Starting with} a thermal equilibrium state of a quantum spin system of $N$ spins, which we call a spin ensemble,
\hk{it is possible to generate} the generalized cat state~\cite{tatsuta2018conversion},
by performing high-resolution, in this case $\Theta(1)$, measurement of magnetization.
However, 
current technology does not allow  
such a high-resolution measurement of the magnetization of
a spin ensemble\hk{.}
Consequently, a more accessible method 
\hk{for creating} a large generalized cat state 
\hk{remains elusive}.
}

\hk{In this study}, we first use the index $q$ to characterize states that are advantageous for quantum sensing,  i.e., beating the SQL although less sensitive than generalized cat states.
We call such  states, which are shown to satisfy $1.5<q<2$,  the SQL-beating catlike states.
We then propose a 
method to create such
SQL-beating states from a thermal equilibrium state of a spin ensemble 
\hk{through} repetitive 
low-resolution
measurement\hk{s} of the total magnetization, utilizing an ancillary qubit coupled with the spin ensemble. 
We prove that we obtain a SQL-beating catlike state with probability $1$ in the large-$N$ limit.
We also perform numerical simulations to observe the gradual emergence of such a state.

This paper is outlined as follows. In Sec. II we review three important notions: the definition of the index $q$, a recipe to create a generalized cat state via single projective measurement, and the basics of quantum metrology.
Section III discusses our finding on the relation between the sensitivity and the value of $q$, in particular when $1.5<q<2$.
In Sec. IV we propose a protocol to create a SQL-beating catlike state via repetitive measurements, in particular within a spin ensemble in a solid state system read out by a superconducting qubit.
In Sec. V we analyze the final state of the spin ensemble after measurements and clarify the condition of successfully generating a generalized cat state. We also discuss the success probability and evaluate how many measurements are required to approximately obtain the state we want.
Numerical simulations are given in Sec. VI, and we indeed observe the emergence of SQL-beating catlike states.
Section VII summarizes the paper and provides an outlook for future work.

\section{Preliminaries}\label{preliminaries}

\as{We first review the previous works on generalized cat
states and a way of obtaining them from thermal equilibrium states
through a single measurement. We also review
Ramsey-type quantum metrology using highly entangled states.}
Throughout this paper, we take $\hbar=1$.
 
\subsection{Generalized cat states}
First, let us briefly review the concept of generalized cat states, characterized by an index denoted by $q$~\cite{q}. For a comprehensive discussion, refer to Ref.~\cite{tatsuta2019quantum}.

The index $q$ \hk{is utilized to}
detect  macroscopic coherence in
a given quantum state $\hat\rho$, which may be pure or mixed.
It is defined through the relation
\begin{align}
\max\left\{\max_{\hat A}\frac{1}{2}\|[\hat A,[\hat A,\hat \rho]]\|_1, N\right\}=\Theta(N^q),
\end{align}
where $\|\hat X\|_1=\mathrm{Tr}\sqrt{\hat X^\dag \hat X}$ is the trace norm and $\hat A$ is an additive observable, \hk{defined as}
$\hat A=\sum_{l=1}^N\hat a(l)$, with $\hat a(l)$ acting on a single spin \hk{at} site $l$
\footnote{Note that $\hat a(l)$ is allowed 
to be 
different from 
the spatial translation of $\hat a(l')$ ($l' \neq l$), i.e., $\hat{A}$ may not be translation invariant.}.
\hk{By} definition, the index
\ym{satisfies \hk{the} condition of}
$1\leq q\leq 2$.
The trace norm can be expressed as
\begin{align}
&\frac{1}{2}\|[\hat A,[\hat A,\hat \rho]]\|_1
=
\max_{\hat \eta} \mathrm{Tr}\left(\hat\eta[\hat A,[\hat A,\hat \rho]]\right)\\
&=
\max_{\hat \eta} \sum_{A,A',\nu,\nu'}(A-A')^2\braket{A,\nu|\hat\rho|A',\nu'}\braket{A',\nu'|\hat\eta|A,\nu},
\end{align}
where  $A$ 
($\ket{A,\nu}$) are eigenvalues (eigenstates) of $\hat A$, 
i.e., $\hat A\ket{A,\nu}=A\ket{A,\nu}$, 
with $\nu$ 
labeling the degeneracy, and $\hat\eta$ is a projection operator, which satisfies $\hat \eta^2=\hat\eta$.
From this expression, we find that $\frac{1}{2}\|[\hat A,[\hat A,\hat \rho]]\|_1$ indicates macroscopic coherence between two macroscopically distinct states and that $q$ can be interpreted as a measure to quantify how close the state is to the cat state.
For example, \hk{it is evident} that macroscopic coherence, i.e., $\braket{A,\nu|\hat\rho|A',\nu'}$ with $|A-A'|=\Theta(N)$ [or $(A-A')^2=\Theta(N^2)$], has a \hk{substantial} weight  in the trace norm of the double commutator if the state has $q=2$.
Then we can conclude that
a state $\hat \rho$ with $q=2$ contains a
substantial amount of
superposition of macroscopically distinct states. 

In addition to the index $q$, we introduce a term ``a generalized cat state.''  We call $\hat\rho$ a generalized cat state of $\hat A$
if
there exists a projection operator $\hat\eta$ such that
\begin{align}
 \mathrm{Tr}(\hat \eta[\hat A,[\hat A,\hat \rho]])=\Theta(N^2)
\end{align}
for a given $\hat A$. From this definition, a generalized cat state has $q=2$.
Conversely, each state with $q=2$ is regarded as a generalized cat state of some observable $\hat{A}$.
As detailed in Sec.~ \ref{ramseyreview}, the generalized cat state exhibits a significant advantage, i.e., beating the SQL, when used in metrology.

\subsection{From a thermal equilibrium state to a generalized cat state}\label{mamineko}
In this section 
\ym{we review a protocol to obtain a generalized cat state from a thermal equilibrium state of a spin ensemble
using 
high-resolution
measurement\hk{s}.}
\hk{We} define $\hat S_\alpha=\sum_{i=1}^N\hat \sigma_\alpha(i)$ with 
\textcolor{black}{$\alpha=x,y,z$}, where $\hat\sigma_\alpha$ 
\hk{denotes} a Pauli operator.

In \cite{tatsuta2018conversion} it was proven
that a single 
\ym{ideal} measurement \ym{of the total magnetization}
can 
\hk{transform} a thermal equilibrium state
into a generalized cat state.
This protocol 
\hk{involves} the following steps.
\hk{
\begin{enumerate}
    \item Prepare the spins in a thermal equilibrium state \as{in} a magnetic field along the $z$ axis, represented by \as{$h$}, at a temperature  $1/\beta$. 
    The Hamiltonian is $\hat H_{\mathrm{P}}=-\omega_\mathrm{P}\hat S_z$, 
    \as{where $\omega_{\mathrm P} = h$.}
    The initial state is then given by $\hat\rho_{\rm P}=e^{\beta \omega_\mathrm{P}\hat S_z}/Z_{\mathrm{P}}$, where $Z_{\mathrm{P}}=\mathrm{Tr}(e^{\beta \omega_\mathrm{P}\hat S_z})$.
    Here the subscript P stands for the phosphorus donor electrons, which we consider as a spin ensemble to generate generalized cat states in Sec.~\ref{setupspinens}.
    \item Perform a projection measurement onto the $\hat S_x=M$ subspace. The post measurement state is expressed as $\hat\rho_{\mathrm{PM}}=\hat{\mathcal{P}}(M)e^{\beta \omega_\mathrm{P}\hat S_z}\hat{\mathcal{P}}(M)/Z_{\mathrm{PM}}$, where $\hat{\mathcal{P}}(M)$ denotes the projection onto the $\hat S_x=M$ subspace and $Z_{\mathrm{PM}}=\mathrm{Tr}(\hat{\mathcal{P}}(M)e^{\beta \omega_\mathrm{P}\hat S_z})$.
\end{enumerate}
}
\hk{It can be shown that}
\begin{align}
\mathrm{Tr}(\hat{\mathcal{P}}(M)[\hat S_z,[\hat S_z,\hat\rho_{\mathrm{PM}}]])=(N^2-M^2)\tanh^2(\beta \omega_\mathrm{P})+2N,\label{maminekotheory}
\end{align}
\hk{indicating} that $\hat\rho_{\mathrm{PM}}$ is a generalized cat state of $\hat S_z$ when $\beta \omega_\mathrm{P}=\Theta(N^0)>0$ and $M\neq \pm N+o(N)$.
Note that this value may be smaller than $\frac{1}{2}\|[\hat S_z,[\hat S_z,\hat\rho_{\mathrm{PM}}]]\|_1$, since taking the trace norm $\|\hat X\|_1$ of the Hermitian $\hat X$
corresponds to applying a projection operator $\hat \eta_M$ that maximizes $2\mathrm{Tr}(\hat\eta_M \hat X)$.
Interestingly, for finite temperature $1/\beta$,
this generalized cat state 
\hk{exhibits} an exponentially small purity, i.e., 
$\mathrm{Tr}(\hat\rho_{\mathrm{PM}}^2)\leq \exp(-\Theta(N))$, due to the highly mixed nature of the premeasurement Gibbs state.
\hk{This explanation pertains to} a simple case 
\hk{with} no interaction\hk{s} between spins.
However, 
it was also proven in \cite{tatsuta2018conversion} that a 
generalized cat state of $\hat S_z$ \hk{can be obtained} even 
\hk{in} the presence of 
interaction
\ym{ between the spins}.

\hk{A significant challenge} of this scheme 
\hk{lies in the requirement for high-resolution} measurements. 
\hk{If} the precision 
of the projection $\hat{\mathcal{P}}(M)$\hk{, equivalent to the minimum resolvable number of spins,} is $\Theta(\sqrt{N})$, the conversion \hk{to a} generalized cat \hk{state} occurs with a probability of $\exp(-\Theta(N))$.  
To 
\hk{achieve a} generalized cat state with a 
probability 
that 
\hk{remains constant as} $N$ increases,
the projection $\hat{\mathcal{P}}(M)$ 
\hk{must be performed with a precision of} $\Theta(1)$\hk{.}
\hk{This level of precision} is experimentally 
\hk{demanding} with a single readout.
\textcolor{black}{In this paper we 
\as{propose to overcome} this difficulty by introducing repetitive \as{measurements}.}

\subsection{Ramsey-type quantum metrology}\label{ramseyreview}
Next let us briefly review Ramsey-type quantum metrology \cite{degen2017quantum} and 
\ym{
\hk{elucidate} how we can 
\hk{utilize} our highly entangled state.}
We adopt this as the basic strategy in our protocol.
\hk{Assume} \ym{that the Hamiltonian 
\hk{describing} the interaction with the magnetic field is $\hat H=-\omega \hat S_z$.}
In a Ramsey-type measurement protocol, 
a sensor state $\hat\rho$ is exposed to the target field for \hk{a} time $t_{\rm int}$, 
resulting in the state evolution
\begin{align}
\hat\rho(t_{\rm int})
&=e^{i \omega t_{\rm int} \hat S_z}\hat\rho e^{-i \omega t_{\rm int} \hat S_z}.
\end{align}
Then a projective measurement is performed. This projection operator can be chosen to optimize the sensitivity.
The probability $P$ of the projection $\hat{\eta}$ is 
given by
\begin{align}
P=\mathrm{Tr}(\hat \eta \hat\rho(t_{\rm int})).
\end{align}
\hk{This protocol is repeated} $T/t_{\rm int}\gg 1$ times, where $T$ is the total measurement time. 
\ym{Here we assume that the state preparation and readout times are 
\hk{negligible compared to} $t_{\rm int}$.}
From the measurement 
\hk{outcomes,} the parameter $\omega$ \hk{can be estimated} with 
\hk{an} uncertainty \cite{huelga1997improvement}
\begin{align}
    \delta\omega = \frac{\sqrt{P(1-P)}}{\left|\frac{dP}{d\omega}\right|}\frac{1}{\sqrt{T/t_{\rm int}}}.
\end{align}
\ym{The inverse of the uncertainty 
\hk{defines} the sensitivity.} 
If the initial state is separable (for which $q=1$), then
\begin{align}
    \delta \omega = \Theta(N^{-1/2}),
\end{align}
\ym{which is 
\hk{known as} the standard quantum limit}.
\hk{Conversely,} 
a state with $q=2$ 
provides us with
\begin{align}
    \delta \omega = \Theta(N^{-1}),
\end{align}
\ym{which is 
\hk{referred to as} the Heisenberg-limited scaling}~\cite{huelga1997improvement}.
This enhancement of sensitivity in scaling is crucial in quantum metrology.

\hk{The} generalized cat state of $\hat A$ \hk{can be employed} to estimate a parameter $\omega$ coupled to $\hat A$, where
the state evolves \hk{under} an interaction Hamiltonian $\hat H=\omega \hat{A}$. 
It has been demonstrated that the target parameter $\omega$ 
in this setup can be estimated at 
 the ultimate scaling sensitivity\hk{:} the Heisenberg-limited scaling $\delta\omega =\Theta(N^{-1})$ in the absence of noise and the Zeno limit $\delta\omega=\Theta(N^{-3/4})$ in the presence of \ym{time-inhomogeneous} dephasing~\cite{tatsuta2019quantum}.
\hk{Additionally}, the method in~\cite{tatsuta2019quantum} %\ymdel{also} 
provides 
a detailed 
\ym{prescription of how to achieve}
this sensitivity 
\hk{using} the generalized cat states.
More specifically, the Ramsey-type protocol is 
utilized, where the generalized cat state is prepared as a probe state,
exposed to the target field, and subsequently measured by
the projection $\hat\eta$.

\hk{It is important to note} that the quantum Fisher information offers the highest sensitivity 
\hk{attainable} with a given state 
\hk{via the} Cram\'{e}r-Rao bound~\cite{paris2009quantum},  assuming an optimal positive-operator-valued measure (POVM)
is chosen.
However, finding the optimal POVM is a core mission in quantum metrology \cite{liu2020quantum} and
its physical implementation may be nontrivial.
\ym{
\hk{In contrast}, the approach 
\hk{using} the generalized cat state \hk{as proposed} in \cite{tatsuta2019quantum} is 
\hk{advantageous because it demonstrates practical} experimental implementation.}
\textcolor{black}{Furthermore, we introduce here a feasible procedure to obtain states with high-$q$ value and explore their features both analytically and numerically.}

In the following section we discuss a relation between index $q$ and 
sensitivity, generalizing the above relations.

\section{SQL-beating catlike states}\label{sec:metuc}
Given the preliminaries discussed above, we now show our first important result
that 
states with $q> 1.5$ can 
serve as 
sensors 
surpassing the SQL.
Specifically,
we show that 
\begin{align}
\delta\omega \leq 1/\Theta(N^{q-1}),
\label{eq:domega<Nq-1}
\end{align}
that is, 
the sensitivity beyond the SQL is achievable for $q>1.5$.

Let $\hat{\eta}$ be the projection operator that satisfies the following:

\begin{align}
\frac{1}{2}\|[\hat S_z,\hat\rho]\|_1=
\mathrm{Tr}\left(\hat\eta[\hat S_z,\hat\rho]\right).
\end{align}
We then obtain the inequality
\begin{align}
\frac{1}{2}\|[\hat S_z,[\hat S_z,\hat\rho]]\|_1
\leq 2N\frac{1}{2}\|[\hat S_z,\hat\rho]\|_1
=2N|\mathrm{Tr}(\hat\eta[\hat S_z,\hat\rho])|.
\end{align}
This inequality implies that if $\hat\rho$ is a generalized cat state of $\hat S_z$, then
$\|[\hat S_z,\hat\rho]\|_1=\Theta(N)$.

Using the Baker-Hausdorff formula, we have
\begin{align}
\hat\rho(t) 
&= \sum_{k=0}^\infty \frac{(i\omega t)^k}{k!}[\hat S_z,\hat\rho]_k,
\end{align}
where we define $[\hat S_z,\hat\rho]_0=\hat\rho$
and
 $[\hat S_z,\hat\rho]_k=[\hat S_z,[\hat S_z,\hat\rho]_{k-1}]$.
 
Hence, we have
\begin{align}
\left|\frac{dP}{d\omega}\right|
&=\left|\mathrm{Tr}\left(\hat \eta \sum_{k=1}^\infty \frac{-it_{\rm{int}}(i\omega t_{\rm{int}})^{k-1}}{(k-1)!}[\hat S_z,\hat\rho]_k\right)
\right|\\
&\simeq \left|\mathrm{Tr}\left(\hat \eta  (-it_{\rm{int}})[\hat S_z,\hat\rho]\right)\right|\quad (\omega t_{\rm{int}}N\ll 1)\\
&=t_{\rm{int}}\left|\mathrm{Tr}\left(\hat \eta  [\hat S_z,\hat\rho]\right)
\right|,
\end{align}
where we have used $\omega t_{\rm{int}}N\ll 1$, which is a common assumption in quantum metrology.

Therefore, we obtain the inequality that ensures the advantage of states with $q>1.5$, 
\begin{align}\label{ineq}
&\delta\omega \simeq \frac{\sqrt{P(1-P)}}{|t_{\rm{int}}\mathrm{Tr}(\hat\eta[\hat S_z,\hat \rho])|}
\frac{1}{\sqrt{T/t_{\rm{int}}}}\\
&\leq \frac{1/2}{\frac{1}{2}\|[\hat S_z,\hat\rho]\|_1}
\frac{1}{\sqrt{Tt_{\rm{int}}}}\\
&\leq \frac{N}{\frac{1}{2}\|[\hat S_z,[\hat S_z,\hat\rho]]\|_1}\frac{1}{\sqrt{Tt_{\rm{int}}}}\label{upperdw}\\
&=\frac{1}{\Theta(N^{q-1})}\frac{1}{\sqrt{Tt_{\rm{int}}}},
\end{align}%atode
which gives \eqref{eq:domega<Nq-1}.

This bound is loose when
considering separable states, 
as they provide
a scaling of $\delta\omega=\Theta(N^{-1/2})$, while
 this bound 
gives $\delta \omega \leq \Theta(N^0)$.
However, this bound is 
significant for states with $q>1.5$, 
because  they 
surpass the SQL in scaling.
Therefore, 
not only the states with $q=2$ 
but also states with $1.5<q<2$ are 
\hk{advantageous} for quantum metrology.
We 
\hk{refer to} such 
states as SQL-beating catlike states
and illustrate their emergence as the result of repetitive measurements.

Note that the inequality \eqref{ineq} provides us with the upper bound of  $\delta\omega$, while the Cram\'{e}r-Rao bound provides the lower bound. More specifically, the actual value of uncertainty might be better than  what we calculate through our inequality, while  the Cram\'{e}r-Rao bound indicates the lowest limit  that might be difficult to achieve because of the imperfection of a realistic setup.

\section{Setup of emergence of SQL-beating catlike states via repetitive measurements}
\ym{We 
\hk{present} our protocol 
\hk{for generating} SQL-beating 
catlike states 
\hk{through} repetitive 
measurements. 
\hk{The procedure involves coupling} 
a spin ensemble with an ancillary qubit
and 
\hk{subsequently reading} out the total magnetization of the spins via the projective measurement of the ancillary qubit.
\hk{Given that} a single measurement of the ancillary qubit \hk{yields a binary result, it does not provide direct}
information on the total magnetization of the spins. However, we 
\hk{demonstrate} that
\hk{through repeated measurements}, the thermal equilibrium state 
\ym{is converted into} 
SQL-beating catlike states due to 
measurement backaction.
Although 
\hk{this} scheme can be 
\hk{implemented in various} physical systems, we 
}
\hk{primarily focus on} 
a hybrid system \hk{consisting} of 
an electron spin ensemble and a superconducting flux qubit  (FQ), which often plays a key role in quantum computing.
Our 
\hk{approach aims to generate generalized} cat states 
\hk{within} the electron spin ensembles by measuring the spin magnetization with the FQ. 
In the following, we briefly review each system and 
\hk{subsequently} propose our scheme.

\subsection{Spin ensemble}\label{setupspinens}
First, we introduce a spin ensemble 
\ym{that we convert into} 
generalized cat states.
\hk{Specifically,} 
\hk{we} consider phosphorus (P) donor electrons in a pure ${}^{28}$Si substrate \cite{tyryshkin2012electron}.
\textcolor{black}{We assume that the temperature is low so that all donor electrons are trapped by the donors.}
The P donor electron 
\hk{possesses a} spin \hk{of} $1/2$.
\as{The donors are distributed randomly in the substrate, and we
assume that the donor density is low, 
typically $10^{15}$ cm${}^{-3}$ \cite{tyryshkin2012electron}, 
so that spin-spin interactions are negligible. }
Notably, 
\hk{these spins exhibit} a long coherence time 
\hk{of} around 10 s \cite{tyryshkin2012electron}, 
\hk{making them} ideal for \hk{the creation and manipulation of}
superposition of macroscopically distinct states.

By applying \hk{a} magnetic field \as{$h$ along the $z$ axis}, the two energy levels are split by the Zeeman energy.
The Hamiltonian is described as
\begin{align}
\hat H_{\mathrm{P}} = -\omega_\mathrm{P} \sum_{j=1}^N \hat\sigma_z(j)=:-\omega_\mathrm{P} \hat S_z,
\label{siliconp}
\end{align}
where $\omega_{\mathrm P} = h$.
This $\omega_\mathrm{P}$ is $\Theta(1)$ and known, unlike the target parameter $\omega$ in Sec.~\ref{ramseyreview}.
\as{The spin ensemble is initially prepared in the canonical Gibbs state
with the inverse temperature $\beta$ \footnote{\as{This preparation is possible 
by, for example, just waiting a long time that is longer than the thermal 
relaxation time, which is much longer than the coherence time.}},
\begin{align}
\hat \rho_{\rm P}(0)=e^{-\beta \hat H_{\rm P}}/\mathrm{Tr}(e^{-\beta \hat H_{\rm P}}),
\label{eq:initialstate}
\end{align}
which has the total spin $\langle \hat S_z \rangle = \Theta(N)$ parallel to $h$.
We will manipulate this state by measuring $\hat S_x$, 
the total spin component perpendicular to $h$, 
with the superconducting flux qubit, which we review in the following subsection.}

\subsection{Superconducting flux qubit}

A superconducting FQ is an artificial two-level system \cite{chiorescu2003coherent,lupacscu2004nondestructive,chiorescu2004coherent,majer2005spectroscopy}, often used in quantum computing.
The FQ 
\hk{comprises}
a superconducting loop  with several (\hk{typically} three) Josephson junctions, operating at \hk{temperatures in the} tens of %\ymdel{mK}
\hk{millikelvin}.
Within the loop, \hk{a} persistent current $I_q$ flows \hk{either} clockwise or counterclockwise. 
The clockwise current corresponds to a state $\ket{R}$, while the counterclockwise current corresponds to 
\hk{the} state  $\ket{L}$.
\hk{By defining} $\hat{\Sigma}_3:=\ket{R}\bra{R}-\ket{L}\bra{L}$ and 
$\hat{\Sigma}_1:=\ket{R}\bra{L}+\ket{L}\bra{R}$,
the Hamiltonian of the FQ \hk{can be expressed}
as
\begin{align}
\hat H_{\rm FQ}= \epsilon \hat \Sigma_3 + \Delta \hat \Sigma_1 \simeq \epsilon \hat \Sigma_3,
\end{align}
where $\epsilon=2I_q(\Phi-\Phi_0/2)$ is an energy bias, 
with $\Phi$ 
\hk{representing the} magnetic flux penetrating 
\hk{the} loop of the FQ and $\Phi_0$ denoting a flux quantum.
Also, $\Delta$ is \hk{the} tunneling energy,
\as{and we assume $\Delta \ll \epsilon$.
Although $\Delta$ is small, it controls the persistent current of the flux qubit \cite{paauw2009tuning,zhu2010coherent,lambert2016superradiance} and will therefore play a crucial role in our scheme.
We will take its effect into account whenever necessary, 
while otherwise we approximate $\hat H_{\rm FQ}$ as $\epsilon \hat \Sigma_3$.}

A flux qubit can be 
\hk{utilized} as a sensitive magnetometer \hk{for} detecting the penetrating magnetic flux $\Phi$ \cite{tanaka2015proposed,bal2012ultrasensitive,toida2019electron,miyanishi2020architecture,budoyo2020electron}.
\hk{To estimate} $\epsilon$, 
which 
\hk{directly translates to estimating}
the external flux
$\Phi$,
the following procedure is employed. 
First, 
the flux qubit \hk{is prepared} in 
\hk{the} state 
\begin{align}
\ket{+}:=(\ket{R}+\ket{L})/\sqrt{2}.
\end{align}
This state can be prepared by applying \hk{a} $\pi/2$ pulse of a resonant microwave \cite{bylander2011noise},
a step referred to as initialization. 
By exposing this state to the target external magnetic field corresponding to $\epsilon$  for a \hk{duration} 
$t$, we obtain $(e^{-i\epsilon t}\ket{R}+\ket{L})/\sqrt{2}$.
\hk{For readout,} we perform a single\hk{-}qubit rotation 
using 
\hk{a} microwave pulse, \hk{followed by measurement using a Josephson bifurcation amplifier \cite{lupacscu2007quantum,lin2013single} or a dispersive readout with a Josephson parametric amplifier.}
\ym{This \hk{process} corresponds to a projective measurement 
\hk{in the} $\hat\Sigma_2:=i\ket{L}\bra{R}-i\ket{R}\bra{L}$ basis.}

\hk{Each} measurement \hk{yields} either $+1$ or $-1$ 
from the readout apparatus of the FQ\hk{,}
with a probability  $P=\cos^2(\epsilon t/2)$ 
for $+1$ 
\hk{and $1-P=\sin^2(\epsilon t/2)$ for $(-1)$}.
By repeating 
\hk{this} procedure, we can accurately  estimate $P$, 
\hk{thereby reducing} the uncertainty 
\hk{in the estimation} of $\epsilon$.
According to the central-limit theorem, the uncertainty $\delta \epsilon$ depends on  the square root of the number of repetitions $m$:
\begin{align}
\delta \epsilon \propto \frac{1}{\sqrt{m}}.
\end{align}
In this paper we increase the $m$ repetitive measurements 
\hk{with} a FQ to convert a thermal equilibrium state into a generalized cat state.

\subsection{Hybrid system}
\hk{We now} introduce the hybrid system 
\hk{comprising} $N$ electron spins and a FQ \cite{marcos2010coupling,twamley2010superconducting,zhu2011coherent,saito2013towards,zhu2014observation}, 
where 
the FQ \hk{serves} as an ancillary qubit to
\hk{repetitively measure the total magnetization of the spin ensemble.}

\begin{figure}[H]
     \centering
     \includegraphics[keepaspectratio, scale=0.5]{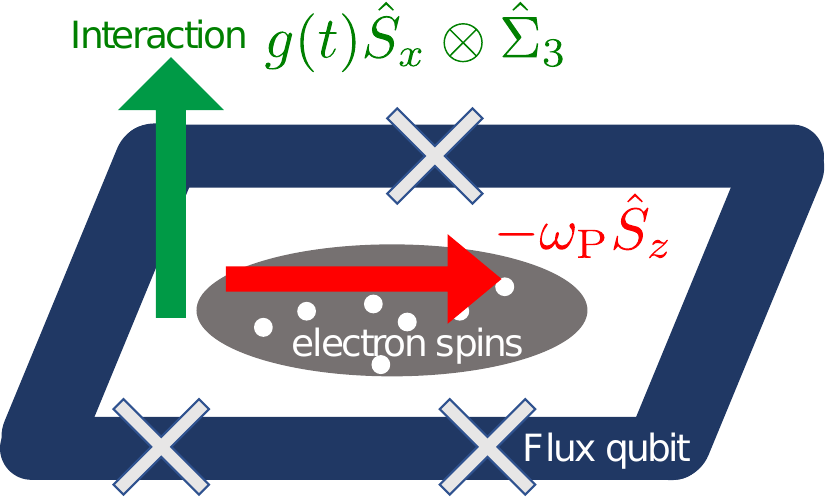}
         \caption{Schematic of the spin ensemble and the FQ. 
         \hk{The} electron spins (white dots) \hk{are situated} on the substrate (gray oval), which is placed within the loop of the FQ (navy parallelogram). 
         \hk{An external} magnetic field (red arrow) is applied along the $z$ axis to induce the Zeeman energy $-\omega_{\rm P}\hat{S}_z$.
         The interaction between the spin ensemble and the FQ is represented by a green arrow.
         }
    \label{ketugou4}
\end{figure}

As illustrated in Fig.~\ref{ketugou4}, an ensemble of donor electron spins in silicon is placed 
inside the loop of the
FQ.
We apply \hk{a} magnetic field parallel to the plane of the FQ loop
\hk{corresponding} to 
\hk{the application of} $\omega_{\mathrm P}\hat S_z$ \hk{as described} in Sec.~\ref{mamineko}.
We assume that the spin ensemble's $x$ axis is orthogonal to the 
\hk{plane of the loop}.
The FQ has two basis states 
\hk{corresponding} to the clockwise and counterclockwise currents. 
These currents generate magnetic fields that shift the Zeeman energy of the elec\hk{t}ron spin in the $x$ direction. 
\hk{This state-dependent} energy shift induces a magnetic interaction between electron spins and the FQ, 
\hk{described by the interaction Hamiltonian}
\begin{align}
\hat H_{\rm int}= g(t)\hat S_x\otimes \hat \Sigma_3,
\end{align}
where $g(t)$ denotes the coupling strength
between them.

The total Hamiltonian \hk{of the system} is
\begin{align}
\hat H =\hat H_{\rm FQ} + \hat H_{\rm P} + \hat H_{\rm int}.
\end{align}
If the interaction induces a frequency shift on the FQ, we 
%\ymdel{could}
\hk{can} use the FQ to detect the magnetization of the spin ensemble by performing the Ramsey\hk{-}type measurements on the FQ. 
\as{We take $h=\omega_\mathrm{P}$ large enough such that  
\begin{align}
|g(t)| \ll \omega_{\rm P}    
\label{eq:g_is_small}
\end{align}
is satisfied.
Under this condition, a dc (time-independent) term in $g(t)$  
is unimportant because it will be dropped due to a rotating wave approximation.
By contrast, an ac (time-dependent) term in $g(t)$
can induce significant resonant effects if 
its frequency is approximately equal to $\omega_{\rm P}$. }
Therefore, we consider \as{the} time-dependent coupling 
\textcolor{black}{
\begin{align}
g(t)=g\cos(2\omega_{\rm P} t)
=g\frac{e^{2i\omega_{\rm P}t}+e^{-2i\omega_{\rm P}t}}{2}.
\label{eq:g(t)}
\end{align}
}
We can 
\hk{achieve} such 
modulation of the coupling strength by 
\hk{altering} the tunneling energy $\Delta$ of the FQ 
\hk{since} the persistent current of the FQ depends on the tunneling energy \cite{lambert2016superradiance}.
\ym{Alte\hk{r}natively, 
\hk{with a time-independent coupling strength}, we could perform many $\pi$ pulses on the FQ 
\hk{sequentially}, 
\hk{akin to} dynamical decoupling \cite{bylander2011noise}, 
\hk{causing} the effective coupling strength 
\hk{to oscillate temporarily}.
In this paper, we focus on the former strategy.}

\as{Because of Eqs.~(\ref{eq:g_is_small}) and (\ref{eq:g(t)}), 
the rotating-wave approximation works well.}
\hk{In the rotating frame defined by} 
\textcolor{black}{$\hat V=\exp(-i\omega_{\rm P}t\hat S_z+i\epsilon \hat \Sigma_3 t)$}\hk{,} we have
\textcolor{black}{
\begin{align}
\hat H_R
&=\hat V \hat H \hat V^\dag -i\hat V\frac{d\hat V^\dag}{dt}
\\
&=
g(t)\frac{\hat S_+e^{-2i\omega_{\rm P} t}+\hat S_-e^{2i\omega_{\rm P} t}}{2}\otimes \hat \Sigma_3,
\end{align}
}
where we 
use
$\hat S_x=(\hat S_++\hat S_-)/2$.
\hk{Thus,} the Hamiltonian in the rotating frame becomes
\begin{align}
\hat H_R
\simeq \frac{g}{2}\hat S_x\otimes \hat \Sigma_3.
\end{align}
This Hamiltonian indicates that the spin magnetization along \hk{the} $x$ axis induces 
\hk{a} frequency shift on the FQ, 
\hk{allowing us to} use 
Ramsey measurements on the FQ to detect the spin ensembles.
\ym{Importantly, in this 
\hk{scenario}, we do not need to control the spin ensemble with 
microwave pulses.}

\begin{figure}[H]
 \centering
\includegraphics[keepaspectratio, scale=0.5]{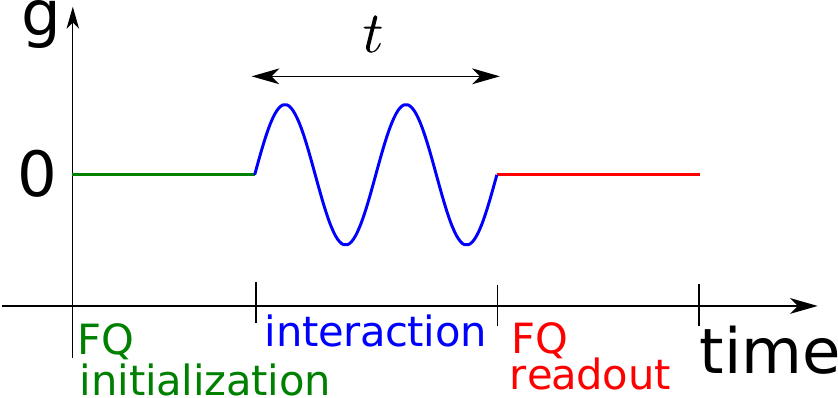}
  \caption{
Schematic of the time dependence of $g(t)$.
\hk{To} induce an interaction between the FQ and \hk{the} spin ensemble, we 
\hk{modulate} the coupling strength to oscillate.
}
\label{ketugou_eng3}
\end{figure}

As depicted in Fig.~\ref{ketugou_eng3}, 
we assume that 
\as{$g(t)=0$}
during \hk{the} initialization and readout of the FQ, \as{whereas it} oscillates as 
\as{Eq.~(\ref{eq:g(t)})}
during interaction.
\ym{Even if the coupling strength has a dc component, we can ignore 
\hk{its} effect,
as mentioned above.}
Below we \hk{analyze} how the state of the spin ensemble \hk{changes} by repeating \hk{these} Ramsey measurements.

\section{Entanglement generation between electron spins via repetitive measurements with the FQ}
In this section we show our second main result that the SQL-beating states  are generated under
the evolution
of
the spin ensemble's state through
a sequence of 
Ramsey-type measurements 
involving
the FQ. 
With the ground state at temperature zero as the initial state of  the spin ensemble,
we obtain a certain  Dicke state
after  infinitely many readout processes in a reasonable parameter regime.
We see that the probability of obtaining a generalized cat state as a final state is asymptotically $1$ for large $N$.

\subsection{State after $m$ measurements}\label{subsec:rhom}
In this section we consider what state we get after $m$ measurements are applied to the spin ensemble.
Let $\hat \rho_{\rm P}(m)$ denote the density matrix of the spin ensemble after \hk{the} %\ymdel{$m$th}
\hk{$m$}{th} measurement by the FQ, 
where \as{$\hat \rho_{\rm P}(0)$ is given by Eq.~(\ref{eq:initialstate}).}
By performing the $(m+1)$th measurement on $\hat \rho_{\rm P}(m)$,
we obtain either of
\begin{align}
\hat \rho_{\rm P}(m+1)=\frac{\hat W_+
\hat\rho_{\rm P} (m)
\hat W_+^\dag
}{{\rm Prob}[\Sigma_2=+1]},
\\
\hat \rho_{\rm P}(m+1)=\frac{\hat W_-
\hat\rho_{\rm P} (m)
\hat W_-^\dag
}
{{\rm Prob}[\Sigma_2=-1]},
\end{align}
\ym{depending on \hk{the} measurement results},
where 
\begin{align}
\hat W_+
&:=
\bra{+_y}
e^{-i\frac{g}{2}\hat S_x\otimes \hat \Sigma_3 t}
\ket{+}
=
\frac{1-i}{\sqrt{2}}
\sin\left(\frac{\pi}{4}+\frac{gt\hat S_x}{2}
\right),
\\
\hat W_-
&:=
\bra{-_y}
e^{-i\frac{g}{2}\hat S_x\otimes \hat \Sigma_3 t}
\ket{+}
=
\frac{1+i}{\sqrt{2}}
\sin\left(\frac{\pi}{4}-\frac{gt\hat S_x}{2}
\right)
\end{align}
are  measurement  operators.
For a detailed derivation, see Appendix \ref{stateaftermeas}.
Here $|\pm _y\rangle $ denotes the eigenvectors of $\hat{\Sigma}_2$ with an eigenvalue of $\pm 1$.
\ym{The probability for the corresponding projection is given 
}
\hk{by}
\begin{align}
{\rm Prob}[\Sigma_2=\pm1]=\mathrm{Tr}(\hat W_{\pm}\hat \rho_{\rm P}(m)\hat W_{\pm}^\dag).
\end{align}

To 
\hk{elucidate} the effect of repetitive measurements, 
\as{we study the zero-temperature limit $\beta \to \infty$ in this section and leave the study of the case of finite temperature to the next section.}
In the zero-temperature limit, the
initial state of the spin ensemble \as{is}
a pure state $\ket{\uparrow}^{\otimes N}$.
Hence we analyze $\ket{\phi_m}$, a pure state after $m$ measurements.
At each readout process, either $\hat W_+$ or $\hat W_-$ 
is probabilistically applied to the state of the spin ensemble. This means 
there are $2^m$ possible trajectories when
\ym{
$m$ measurements \hk{are performed} by the FQ.}
Fortunately, since $\hat W_+$ commutes with $\hat W_-$,
the number of \hk{effective} trajectories \hk{is reduced} and there are only  $m+1$ distinct trajectories to consider.
Let $k$ 
denote 
\hk{the number of times} $\hat W_+$ is applied to the state 
during
 the $m$ measurements. 
For a given $k$, the final state can be expressed as
\begin{align}
\ket{\phi_m}=(\hat W_+)^k(\hat W_-)^{m-k}\ket{\uparrow}^{\otimes N}/(\mathrm{norm}),\label{phim}
\end{align}
where 
\begin{align}
(\mathrm{norm})^2=\bra{\uparrow}^{\otimes N}(\hat W_+^\dag\hat W_+)^k(\hat W_-^\dag \hat W_-)^{m-k}\ket{\uparrow}^{\otimes N}.
\end{align}
\hk{Assuming} 
\begin{align}
gtN<\pi/2   
\end{align}
for simplicity for a while,
we can 
{prove}
 that 
for $m\gg1$,
$(\hat W_+)^k(\hat W_-)^{m-k}$ 
\hk{attains} the eigenvalue with the largest modulus
for eigenstates of $\hat{S}_x$, when the corresponding eigenvalue $S_x$ becomes the closest to $(gt)^{-1}\arcsin\left(2k/m-1\right)$.
\hk{Details are provided in Appendix \ref{appB}.}

\ym{Importantly, it is known that 
if 
an
operator $\hat K$ \hk{is applied}
\ym{infinitely many times}
to the system,
the system converges to 
the eigenstate of $\hat K$ 
corresponding to the eigenvalue with the largest modulus,
provided 
it is unique~\cite{nakazato2003purification,stockton2004deterministic,dasari2022anti}. 
Furthermore, 
in our case,  we obtain $(\hat W_+)^k(\hat W_-)^{m-k}=((\hat W_+)^{\alpha }(\hat W_-)^{(1-\alpha)})^m$ for $k=\alpha m$ where $0\leq\alpha \leq 1$. It is worth mentioning that 
$\hat{K}=(\hat W_+)^{\alpha }(\hat W_-)^{(1-\alpha)}$ 
and
$(\hat W_+)^k(\hat W_-)^{m-k}$
take the largest eigenvalue with the same eigenstate.
By 
\hk{applying} this fact to our case, we obtain}
\begin{align}
\ket{\phi_m} \simeq \left| S_x=(gt)^{-1}\arcsin\left(\frac{2k}{m}-1\right)\right\rangle,\label{noninteger}
\end{align}
\ym{for large $m$, if we assume that $(gt)^{-1}\arcsin\left(\frac{2k}{m}-1\right)$ is an integer.}
Here we define 
\begin{align}
\ket{S_x=\theta}:=\ket{D_N^{(\theta)}}, 
\end{align}
\hk{where}
$\ket{D_N^{(\theta)}}$ is a Dicke state given by
\begin{align}
\ket{D_N^{(\theta)}}=\sqrt{\binom{N}{\frac{N+\theta}{2}}^{-1}}\sum_{\sigma\in \mathcal{S}_N}
\mathcal{P}_\sigma\left(\ket{+}^{\otimes (N+\theta)/2}\ket{-}^{\otimes (N-\theta)/2}\right), \label{dicke}
\end{align}
\hk{with} $\mathcal{P}_\sigma$ 
\hk{denoting} the permutation of spins and the summation 
taken over all 
different permutations in the permutation group $\mathcal{S}_N$
\cite{dicke1954coherence,stockton2003characterizing,kiesel2010operational}.

For an illustration,
consider the 
example
\hk{where} $k=m/2$ for an even number $m$.
\hk{An} operator
$\hat W_+\hat W_-=\cos(gt\hat S_x)/2$ 
is applied $m/2$ times.
In the region $-\pi/2< x<\pi/2$, the function $\cos(x)$ 
\hk{attains its} maximum value \hk{of} $1$ at $x=0$.
Therefore\hk{,} we obtain 
\begin{align}
\ket{\phi_m}\rightarrow \ket{S_x=0}\label{k=m/2}
\end{align}
 in the limit of $m\rightarrow \infty$.

Note that $S_x$ takes 
\hk{only integer values}, while $(gt)^{-1}\arcsin\left(\frac{2k}{m}-1\right)$ is not necessarily an integer.
This 
\as{necessitates refining the 
formula (\ref{noninteger}) as follows.}
In the limit of $m\rightarrow \infty$, the final state converges to 
\begin{align}
\lim _{m\rightarrow \infty}\ket{\phi_m}= \ket{S_x=L},
\end{align}
where $L$ is the integer closest to $(gt)^{-1}\arcsin\left(2k/m-1\right)$.
If $(gt)^{-1}\arcsin\left(2k/m-1\right)$ 
\ym{is}
\hk{a half-integer}, 
\ym{there are two possible states 
\hk{with} eigenvalues whose modulus is the largest. In this case, the final state depends on 
\hk{the initial state's weight} about these two states. }
This idea can be applied also in the next paragraph.

In \ym{ actual}
experiments, 
\ym{it is not always possible to satisfy the condition of $gtN< \pi/2$.}
In a region $gtN\geq \pi/2$,
the operator
$(\hat W_+)^k(\hat W_-)^{m-k}$
takes the eigenvalue with the largest modulus when $S_x$ takes 
\as{either of the following values:}
% the values either
\begin{align}
\theta_1(n):=\left(\arcsin\left(\frac{2k}{m}-1\right)+2n\pi\right)/(gt),\\  
\theta_2(n'):=\left(\pi-\arcsin\left(\frac{2k}{m}-1\right)+2n'\pi\right)/(gt).
\end{align}
Here, $n$ and $n'$ are integers that satisfy 
\begin{align}
-gtN &\leq
\arcsin(\frac{2k}{m}-1)+2n\pi
 \leq gtN,
 \\
-gtN &\leq
\pi-\arcsin(\frac{2k}{m}-1)+2n'\pi
\leq gtN,
\end{align}
where we use
$-N \leq  S_x \leq N$.
As we have discussed in the preceding paragraph, the final state should be $\ket{S_x=L}$, where $L$ is the integer closest to the largest (in magnitude) eigenvalue of $(\hat W_+)^k(\hat W_-)^{m-k}$, among $\theta_1(n)$ and $\theta_2(n')$ with multiple candidates of $n$ and $n'$ (we assume that such $L$ is unique for simplicity).
This implies that the integer $L$ is taken as
\begin{align}
L= \left\{
\begin{array}{ll}
\theta_1(n) & (\min_{n\in \mathbb{Z}}
h_1(n) \leq \min_{n'\in \mathbb{Z}}
h_2(n'))\\
\theta_2(n') & (\min_{n\in \mathbb{Z}}
h_1(n) > \min_{n'\in \mathbb{Z}}
h_2(n'))
\end{array}
\right.
\end{align}
where 
\begin{align}
&h_1(n):=\left|
\theta_1(n)-\mathrm{round}(\theta_1(n))
\right|,\\
&h_2(n'):=
\left|
\theta_2(n)-\mathrm{round}(\theta_2(n))
\right|.
\end{align}
Here $\mathrm{round}(x)$ is the integer closest to $x$.

\subsection{Variance of $\hat S_z$ in $\hat\rho_{\mathrm{P}}(m)$}

Next we \hk{examine} whether 
\hk{these} postmeasurement states are generalized cat states.
As we have clarified in the preceding section, the final state, i.e., 
$\hat\rho_{\mathrm{P}}(m)$ with $m\rightarrow \infty$,
is a pure state when the initial state is a pure state $\ket{\uparrow}^{\otimes N}$.
If 
\hk{the} variance of an additive observable is 
\hk{of the} order of $N^2$, such a pure state is 
a generalized cat state
\cite{q,tatsuta2019quantum}.

Calculating the variance of $\hat S_z$ for $\ket{S_x=\xi}$, 
we obtain
\begin{align}
&\braket{ S_x=\xi|\hat S_z^2| S_x=\xi}
-
\braket{ S_x=\xi|\hat S_z| S_x=\xi}^2
\nonumber\\
&=
\frac{N^2-\xi^2}{2}+N.\label{nekojouken}
\end{align}
This shows that, 
as long as there exists a positive $N$-independent constant $\delta$ ($< 1$) such that $|\xi /N|< 1-\delta$ is satisfied,
 $\ket{S_x=\xi}$ is a generalized cat state.
In the case $gtN<\pi/2$,
the final state
$\ket{S_x\simeq(gt)^{-1}\arcsin((2k-m)/m)}$ 
is a generalized cat state if 
there exists a positive constant $\delta'$ ($<1$) such that
$|(Ngt)^{-1}\arcsin((2k-m)/m) |< 1-\delta'$ is satisfied,
assuming for simplicity that $L$ in the preceding section is uniquely determined.
Intuitively, this can be explained as follows:
Generalized cat states are produced by the $m$ measurements
when $k$ is far enough from $0$ and $m$ 
so that $(gt)^{-1}\arcsin((2k-m)/m)$ is not as large as $N$.

\subsection{Probability of trajectories }
We investigate 
the success probability of creating a generalized cat state
\hk{when} the initial state is $\ket{\uparrow}^{\otimes N}$. 

\begin{figure}[H]
 \centering
\includegraphics[keepaspectratio, scale=0.4]{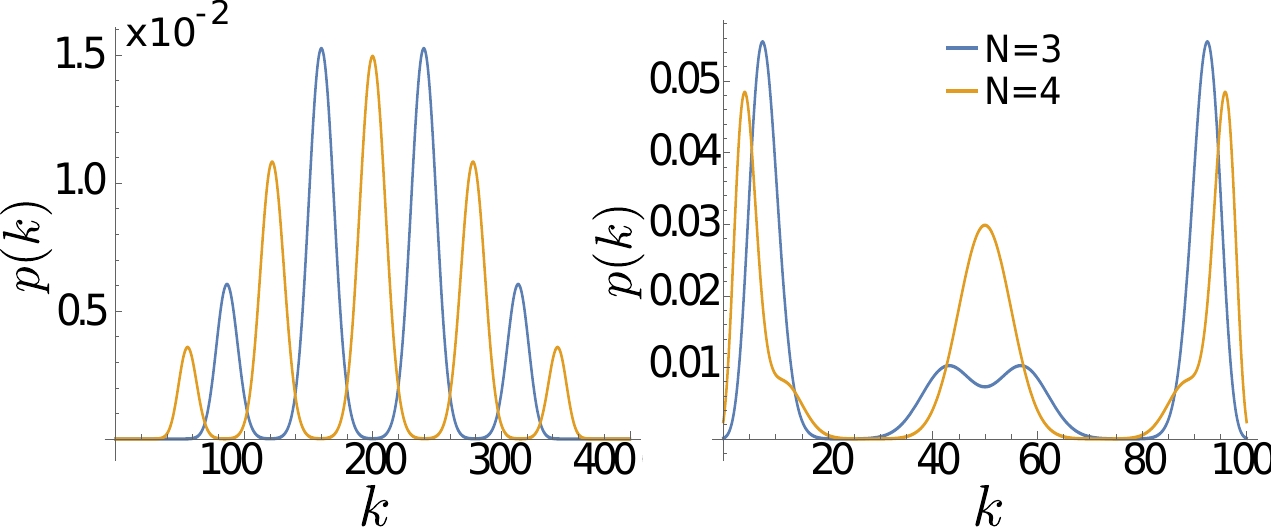}
  \caption{
Plot of the probability $p(k)$ of obtaining the trajectories in which $\hat W_+$ is applied $k$ times\hk{, as} described in  Eq.~(\ref{withoutappr})\hk{,} against $k$.
In (a) \hk{the} parameters are $gt=0.2$, $m=400$, and $N=3$ (blue) and $N=4$ (yellow), satisfying $gtN< \pi/2$. 
In this case, $p(k)$ is approximately given by the sum of the sharp normal-distribution-like peaks, whose heights become larger when they are closer to $k=m/2$.
In (b) \hk{the} parameters are $gt=1.0$, $m=100$, and $N=3$ (blue) and $N=4$ (yellow), satisfying $gtN\geq \pi/2$. 
}
\label{trapro}
\end{figure}

Let
\hk{denote} the probability of obtaining the trajectories in which $\hat W_+$ is applied $k$ times.
Since
there are $\binom{m}{k}$ trajectories
for a given $k$,
the probability $p(k)$ is 
calculated
as (see Appendix \ref{appProb} for derivation)
\begin{align}
&p(k)
\binom{m}{k}
\bra{\uparrow}^{\otimes N}
\left(\hat W_+^\dag\hat W_+\right)^k
\left(\hat W_-^\dag\hat W_-\right)^{m-k}
\ket{\uparrow}^{\otimes N}
\\
&=
\binom{m}{k}
\frac{1}{2^N}
\sum_{r=0}^N
\binom{N}{r}
\nonumber\\
&\times
\left(\frac{1+\sin(gt(2r-N))}{2}\right)^k
\left(\frac{1-\sin(gt(2r-N))}{2}\right)^{m-k}.
\label{withoutappr}
\end{align}
This $p(k)$ 
\hk{exhibits markedly} different behaviors for 
\hk{various} parameters (Fig.~\ref{trapro}).
For 
$gtN< \pi/2$, there are two highest peaks near $k=m/2$ 
\hk{for} odd $N$, 
\hk{whereas} there is 
\hk{a single} highest peak at $k=m/2$ %\ymdel{with}
\hk{for} even $N$.
\hk{Additionally}, the probabilities 
$p(0)$ and $p(m)$ are small.

Notably, we can prove that success probability of generating a generalized cat state is sufficiently large.
In particular, we can show that the probability approaches $1$ as
\begin{align}
\lim_{N\rightarrow\infty}\lim_{m\rightarrow\infty} 
\sideset{}{'}{\sum}_{k} p(k) = 1
\quad \mbox{when } gtN<\pi/2,
\label{eq:sucprob_to_1}
\end{align}
where $\sum_{k}'$ denotes the sum taken over $k$ such that the corresponding stationary states are the generalized cat states.
In fact, in the limit $m\rightarrow \infty$, $p(k)$ can be regarded as a sum of the sharp normal-distribution peaks (as in Fig.~\ref{trapro}(left)).
Then, taking $N\rightarrow \infty $ leads to the concentration of the peaks towards $k\simeq m/2$, resulting in  $\sum_k' p(k)\rightarrow 1$.
For the detailed proof, see Appendix \ref{appProb1}.
This is intuitively because we obtain a generalized cat state unless $k$ is close to $0$ or $m$ when $gtN<\pi/2$, which is unlikely as already speculated from  Fig.~\ref{trapro}(a).

For 
$gtN\geq \pi/2$, $p(k)$ exhibits
\ym{a different behavior [see Fig.~\ref{trapro}(b)].}
We 
\hk{observe} that $p(0)$ and $p(m)$ have relatively large values.
\ym{However,} 
we still obtain generalized cat states \ym{with a reasonable success probability}\hk{,}
as we numerically find in the next section.

\subsection{
Required number of measurements}
As explained in Sec.~\ref{subsec:rhom}, an infinite number of measurements convert the initial state to $\ket{{S}_x=L}$.
In this section we discuss how many measurements are required to approximately obtain this stationary state, assuming the case $gtN<\pi/2$.

For this purpose, we evaluate the spectral gap characterizing the dynamics.
More precisely, we consider the spectrum $\{\lambda_a\}\:(\lambda_1\leq \lambda_2\leq\cdots)$ of an operator
\begin{align}
-\frac{1}{2m}\ln \left(\hat W_+^k\hat W_-^{m-k}\right)\left(\hat W_+^k\hat W_-^{m-k}\right)^\dag
=
-\frac{1}{2}\ln \hat{K}\hat{K}^\dag
\end{align}
with
$\hat{K}=\hat{K}^\dag=(\hat W_+)^\alpha(\hat W_-)^{1-\alpha}$
and $\alpha=k/m$, which is called the Lyapunov spectrum of a trajectory under measurement~\cite{benoist2019invariant,PhysRevLett.134.010410}.
The inverse of the gap of the Lyapunov spectrum, $\lambda_2-\lambda_1$, is known to provide the timescale for relaxation to the final state.
This is because the ratio of weights between the longest-lived decaying mode and the final state is evaluated by $e^{-\lambda_2 m}/e^{-\lambda_1 m}$, 
which becomes approximately $e^{-1}$ when $m \simeq m_\mathrm{relax} := (\lambda_2-\lambda_1)^{-1}$.

Since $\hat W_\pm$ are diagonal in the $\hat S_x$ basis, the above operator is simply given by
\begin{align}
-\ln f(gt\hat{S}_x),
\end{align}
where $f(x):=\sin^\alpha\left(\frac{\pi}{4}+\frac{x}{2}\right)
\sin^{1-\alpha}\left(\frac{\pi}{4}-\frac{x}{2}\right)
$ (see Appendix~\ref{appB}).
In Appendix~\ref{appB} we show that $f(x)$ takes a maximum value when
\begin{align}
x_*=\arcsin\left(2\alpha-1\right),
\end{align}
which leads to Eq.~\eqref{noninteger} and $\lambda_1=-\ln f(x_*)$ if $x_*/gt$ is an integer.
If $ x_*/gt$ is not an integer, we instead have
\begin{align}
\lambda_1=-\ln f(gt L)
\end{align}
with $L=\mathrm{round}(x_{*}/gt)$.
%Without loss of generality, we assume $x_{**}\leq x_*$ in the following.

Next, since $f(x)$ is a concave function taking the maximum at $x=x_*$ (Appendix~\ref{appB}) and ${S}_x$ takes integer values, we can evaluate $\lambda_2$ as
\begin{align}
\lambda_2=-\ln f(gtL+ gt)
\end{align}
or
\begin{align}
\lambda_2=-\ln f(gtL- gt).
\end{align}

Let us first assume that $x_*/gt$ is an integer, i.e., $gtL=x_*$.
Then, assuming $gt\ll 1$, we can evaluate the gap at the leading order of $(gt)^2$ from the form of $f(x)$, finding
\begin{align}
\lambda_2-\lambda_1\simeq \frac{(gt)^2}{4}.
\end{align}
This leads to the relaxation timescale
\begin{align}
m_\mathrm{relax}\simeq \frac{4}{(gt)^2}.
\end{align}
Interestingly, this does not explicitly depend on $\alpha$.

When $x_*/gt$ is not an integer, the gap becomes smaller than $(gt)^2/4$ but remains the order of $(gt)^2$, unless it accidentally vanishes. Therefore, we conclude that the relaxation time for most measurement outcomes becomes the order of $m_\mathrm{relax}\sim (gt)^{-2}$.

\section{Numerical simulations}

In this section, 
\as{we study the case of finite temperature using}
numerical simulations.
\hk{The} initial state of the spin ensemble
\as{is}
\textcolor{black}{a mixed state, i.e.,}
a thermal equilibrium state
\hk{subjected to} a magnetic field $h$ along the $z$ axis.
The quantity that we mainly focus on is 
\begin{align} 
    C_{\mathrm cat}(\hat S_z, \hat\rho_\mathrm{P}(m)):=\frac{1}{2}\|[\hat S_z,[\hat S_z,\hat{\rho}_\mathrm{P}(m)]]\|_1.
    \label{ccat}
\end{align}
If it is $\Theta(N^q)$ with $q> 1.5$, the state is a SQL-beating catlike state, as discussed in Sec.~\ref{sec:metuc}.
\hk{To visualize the emergence of} generalized cat states 
we numerically 
\hk{compute} the value of $ C_{\mathrm cat}(\hat S_z, \hat\rho_\mathrm{P}(m))$
\hk{for} each $m$ up to $m=1000$.
\as{In the following, we take 
$1/\beta=0.1$ and $h=0.5$.}
\hk{The} interaction strength between the FQ and the spin ensemble 
\hk{is set to be} $gt=0.222$.
This is the case where $gtN>\pi/2$, which we did not analytically discuss above. Remarkably, we can see the emergence of SQL-beating catlike state even in this regime.

\subsection{Trajectories for a fixed $N$}
\begin{figure}[H]
 \centering
\includegraphics[keepaspectratio, scale=0.65]{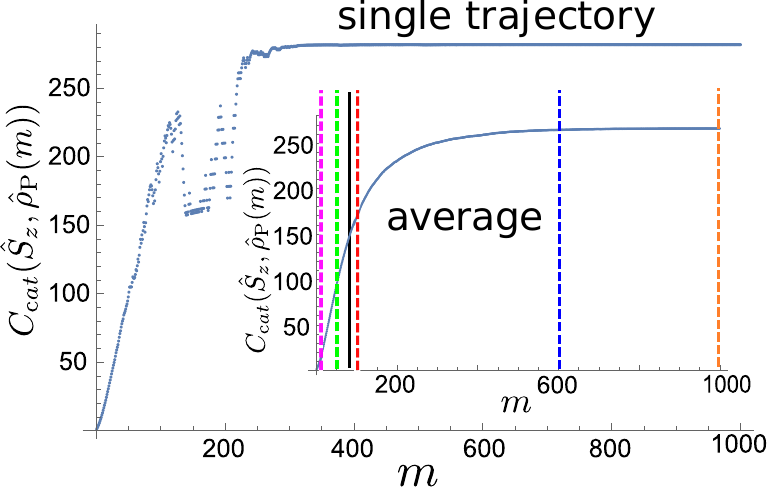}
  \caption{
Visualization of the creation of 
generalized cat states 
\hk{through} repetitive measurements
 with $N=15$ qubits.
 The horizontal axis 
 \hk{represents} the number of measurements with the FQ, 
\hk{while} the vertical axis denotes
 the value of 
 $C_{\mathrm cat}(\hat S_z, \hat\rho_\mathrm{P}(m))$.
\hk{Due to the probabilistic nature of} each measurement, 
the values of 
$C_{\mathrm cat}(\hat S_z, \hat\rho_\mathrm{P}(m))$
exhibit temporal fluctuations within 
\hk{a} single trajectory. 
The inset 
\hk{illustrates} the results 
averaged over \hk{$3000$} runs. 
In the inset, the black vertical line shows $m=82\simeq m_\mathrm{relax}$. Purple, green, red, blue and orange lines in the inset indicate $m=10, 50, 100, 600, 1000$, respectively.
\hk{This average reveals a clear,} gradual increase 
\hk{in} 
$C_{\mathrm cat}(\hat S_z, \hat\rho_\mathrm{P}(m))$.
The initial state is $\exp(-\beta h\hat S_z)/\mathrm{Tr}\exp(-\beta h\hat S_z)$ with 
 $1/\beta=0.1$, \textcolor{black}{$h=0.5$}, and \hk{a} coupling strength \ym{multiplied by the interaction time}
 given by
  $gt=0.222$.
  }
\label{n15}
\end{figure}

Let us 
\hk{examine} a
single trajectory for a  fixed $N\:(=15)$,
\hk{depicted} in  Fig.~\ref{n15}.
In 
\hk{each} trajectory, 
\ym{the measurement 
\hk{outcome} at each $m$ is probabilistically 
\hk{determined},}
and the measurement backaction 
from each 
result 
\hk{influences}  
the value of $C_{\mathrm cat}(\hat S_z, \hat\rho_\mathrm{P}(m))$.
Due to the stochastic nature of 
\hk{these} measurements,
a single trajectory
\hk{exhibits} temporal fluctuations\hk{, albeit with an overall tendency for 
$C_{\mathrm cat}(\hat S_z, \hat\rho_\mathrm{P}(m))$
to increase.} 
By 
\hk{averaging over} $3000$ 
\hk{trajectories}, we observe that 
the value of $C_{\mathrm cat}(\hat S_z, \hat\rho_\mathrm{P}(m))$
\hk{undergoes rapid} changes 
\hk{initially but becomes almost constant around}  $m\sim 600$, as 
\hk{shown} in the inset of  Fig.~\ref{n15}.
Note that at $m_\mathrm{relax}\simeq 4/(gt)^2\simeq 82$, the value of 
$C_{\mathrm cat}(\hat S_z, \hat\rho_\mathrm{P}(m))$
is 147, which is nearly half of the  stationary 
value of $C_{\mathrm cat}(\hat S_z, \hat\rho_\mathrm{P}(m))$,
even though we consider here the case $gtN>\pi/2$.

\subsection{Scaling behavior of 
$C_{\mathrm cat}(\hat S_z, \hat\rho_\mathrm{P}(m))$
}
\begin{figure}[H]
 \centering
\includegraphics[keepaspectratio, scale=0.46]{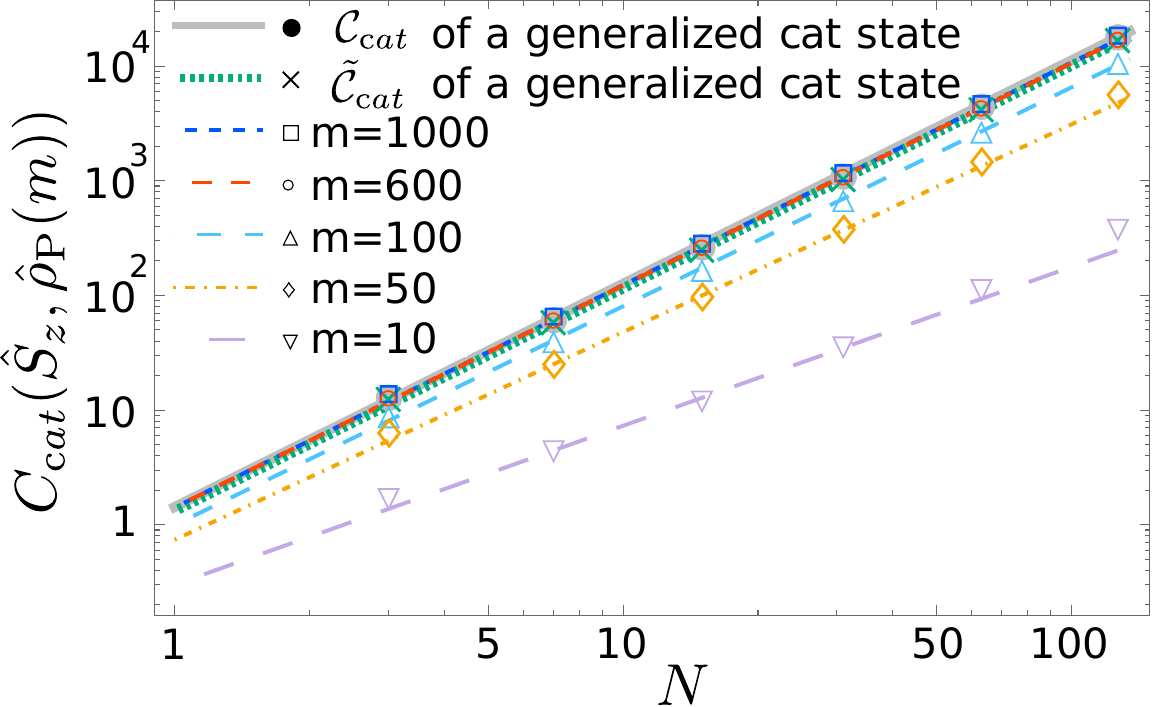}
  \caption{A log-log plot of 
  the value of $C_{\mathrm cat}(\hat S_z, \hat\rho_\mathrm{P}(m))$
  against $N$ for different $m$
  \hk{values, along with} two reference plots. 
From the bottom to the top, \hk{the $m$ values are}
$m=10$ (purple), $50$ (orange), $100$ (cyan), $600$ (red), and $1000$ (blue). 
The initial state is \hk{given by} $\exp(-\beta h\hat S_z)/\mathrm{Tr}\exp(-\beta h\hat S_z)$ with $1/\beta=0.1$, \textcolor{black}{$h=0.5$}, and \hk{a} coupling strength 
\hk{of} $gt=0.222$.
The thick gray line \hk{represents the} reference obtained from Eq.~(\ref{ideal}), which is 
$\mathcal{C}_{\mathrm cat}$
of a generalized cat state.
The green dotted line \hk{represents the} reference value of 
$\tilde{\mathcal{C}}_{\mathrm cat}$
obtained from Eq.~(\ref{maminekoexpect}). 
\hk{Error bars were omitted as} they were smaller than the widths of the dots.
}
\label{katamuki}
\end{figure}

To 
\hk{determine} 
\ym{whether we 
\hk{have successfully created a} generalized cat state}, 
it is
\hk{essential} to 
\hk{examine} the scaling behavior of the 
$C_{\mathrm cat}(\hat S_z, \hat\rho_\mathrm{P}(m))$
\hk{with respect to} $N$
at each $m$.
For $N=3,\,5,\,7,\,15,\,31,\,63,\,127$, we  
average 
$3000$ random trajectories and plot the value of $C_{\mathrm cat}(\hat S_z, \hat\rho_\mathrm{P}(m))$
in Fig.~\ref{katamuki}.
\hk{A} linear fit \hk{is performed} to determine the slope of the line, 
\hk{corresponding} to the index $q$.
As 
\hk{depicted}, the slope increases 
\hk{with} $m$,
but 
\hk{does not show a significant change} after $m\sim 600$.

With a linear fit using the data 
for $N=3,\,5,\,7,\,15,\,31,\,63,\,127$,
we obtain up to $q=1.94$ at $m=1000$, which is sufficiently close to the generalized cat state with $q=2$.
Since $q>1.5$ is achieved already for $m=50$, we can create SQL-beating catlike states by this timescale.

Note that, for reference, Fig.~\ref{katamuki} also shows two additional curves which concern a generalized \ym{cat state} $\hat\rho_{\rm PM}$ obtained by a single
projective measurement $\hat{\mathcal{P}}(M)$ \cite{tatsuta2018conversion}
(see Sec.~\ref{mamineko}), where the initial state is the zero-temperature pure
state $\hat\rho_{\rm pure}=\ket{\uparrow}^{\otimes N}\bra{\uparrow}^{\otimes N}$.
One curve describes 
$C_{\mathrm cat}(\hat S_z, \hat\rho_{\rm PM})$
of the postmeasurement state averaged over outcomes $M=-N,-N+2,\cdots,N$ for the projection measurement,
\begin{align}
\mathcal{C}_{cat}
:= &\sum_{M=-N,N-2,...,N} \frac{1}{2^N}\binom{N}{\frac{N+M}{2}}
\notag\\
&\qquad \qquad \times \frac{\|[\hat S_z,[\hat S_z, \hat{\mathcal{P}}(M)\hat\rho_{\rm pure} \hat{\mathcal{P}}(M)]]\|_1}{2}
\label{ideal}
\end{align}
where $\binom{N}{(N+M)/2}/2^N$ is the corresponding probability. 
If the initial state is a mixed state at finite temperature, the average value of $C_{\mathrm cat}(\hat S_z, \hat\rho_{\rm PM})$ is smaller than $\mathcal{C}_{\rm cat}$.
The second curve 
is associated with
\begin{align}
\tilde{C}_{\mathrm cat}(\hat S_z, \hat\rho_\mathrm{PM}):=\mathrm{Tr}(\hat{\mathcal{P}}(M)[\hat S_z,[\hat S_z,\hat\rho_{\mathrm{PM}}]]),
\end{align}
which provides the analytically tractable lower bound of Eq.~\eqref{ccat}.
Motivated by the intuitive measurement protocol explained in Sec.~\ref{mamineko}, 
$\tilde{\mathcal{C}}_{\mathrm cat}$
is defined as the zero-temperature limit ($\beta\rightarrow\infty$) of Eq.~\eqref{maminekotheory} averaged over all measurement outcomes, i.e.,
\begin{align}
\tilde{\mathcal{C}}_{\rm cat}:=
&\sum_{M=-N,N-2,...,N}\frac{1}{2^N}\binom{N}{\frac{N+M}{2}}\mathrm{Tr}(\hat{\mathcal{P}}(M)[\hat S_z,[\hat S_z,\hat\rho_{\mathrm{PM}}]])
\nonumber \\
&=\sum_{M=-N,N-2,...,N}\frac{1}{2^N}\binom{N}{\frac{N+M}{2}}\left(N^2-M^2+2N\right).\label{maminekoexpect}
\end{align}

As can be seen in Fig.~\ref{katamuki}, the fitted line of the state after $600$ measurements is indistinguishable from $\mathcal{C}_{\rm cat}$ and $\tilde{\mathcal{C}}_{\rm cat}$,
the plots for the generalized cat state generated 
via single projective measurement at zero temperature. % $\hat\rho_{\rm pure}$.
This implies that the repetitive measurements we consider in this paper can indeed replace the single-projective-measurement scheme in Ref \cite{tatsuta2018conversion}, which required considerably high-resolution measurement.
It is also remarkable that our scheme is comparable to the single projective-measurement scheme under zero temperature,
even though we consider nonzero temperature.

\ym{Finally, we note that within this numerical simulation, we cannot
\hk{achieve} $\Theta(N^2)$ scaling of 
$C_{\mathrm cat}(\hat S_z, \hat\rho_\mathrm{P}(m))$
from the fitting.
The possible reasons \hk{for this} are as follows.
First,  we considered the region $gtN>\pi/2$ in the numerical simulation, while we proved \eqref{eq:sucprob_to_1} assuming $gtN<\pi/2$.
Second, unless \hk{$N$ is infinitely large}, 
there 
\hk{will be} a finite contribution from 
\hk{both} $\Theta(N^2)$ 
\hk{and} $\Theta(N)$ \hk{terms} even when the generalized cat states are considered, as can be seen in Eq.~(\ref{maminekotheory}). 
\hk{When} we substitute $N=3,\,5,\,7,\,15,\,31,\,63,\,127$ in Eqs.~(\ref{maminekoexpect}) and~(\ref{ideal})
and perform the linear fitting in the same 
\hk{manner}, we obtain $q=1.95$,
\hk{which is less} than  $2$, even when 
\hk{accounting for} fitting error~\footnote{Note that we get $q=1.995$, almost $q=2$, when we numerically calculate the value of 
$C_{\mathrm cat}(\hat S_z, \hat\rho_\mathrm{PM})$
of a generalized cat state with $N=63$, $127$ and $1027$ and perform a fitting, from which we expect $q\rightarrow 2$ with large $N$ even with our scheme.
}.
\hk{Consequently, it is challenging to obtain}  
$C_{\mathrm cat}(\hat S_z, \hat\rho_\mathrm{P}(m))$
with the scaling $\Theta (N^2)$ from 
numerical simulations up to $N=127$.
\hk{Nonetheless}, our results are still
\hk{significant as they demonstrate the potential of utilizing} this highly entangled state with $q>1.5$ for quantum sensing, as we discussed above.
}

\subsection{Sensitivity of the generated states}
Let us calculate the sensitivity of the states with $q>1.5$ that are generated through the repetitive measurements.
We consider the case where  $N=127$. 
%Let $m$ be the number of the measurements made.
Substituting the value of  
$C_{\mathrm cat}(\hat S_z, \hat\rho_\mathrm{P}(m))$
for Eq.~(\ref{upperdw}), we show 
%we substitute the value of catness for Eq.~(\ref{upperdw}) to calculate
$\delta\omega$ of the following states in Table \ref{kandosuuchi}: the state at $m=50$ with  $q=1.81$, the state at $m=100$ with $q=1.91$, the state at $m=600$ with $q=1.94$, and generalized cat state with  $q=2$ obtained by a single projective measurement.
We also calculate the sensitivity of a separable state using a known formula 
$\delta\omega=\frac{1}{2\sqrt{N}\sqrt{Tt_{\rm{int}}}}$.

\begin{table}[h]
    \centering
    \caption{Comparison of sensitivity of states with various $q$. The uncertainty $\delta\omega$ for states with $q=1, 1.81, 1.91, 1.94, 2$ is calculated. We can see a clear advantage of $q>1.5$.}
    \label{kandosuuchi}
    \begin{tabular}{ccc}
        \hline \hline
        $m$ & $q$ & $\delta\omega$ \\ \hline
         & $1$ (separable state) & $ 4.4\times 10^{-2}/\sqrt{T t_{\rm int}}$ \\ 
$50$ & $1.81$ & $\leq 2.4\times 10^{-2}/\sqrt{T t_{\rm int}}$ \\ 
$100$ & $1.91$ & $\leq 1.2\times 10^{-2}/\sqrt{T t_{\rm int}}$ \\ 
$600$ & $1.94$ & $\leq 7.4\times 10^{-3}/\sqrt{T t_{\rm int}}$ \\ 
 & $2$ (generalized cat state) & $\leq 7.0\times 10^{-3}/\sqrt{T t_{\rm int}}$ \\ \hline\hline
    \end{tabular}
\end{table}

From this table it is clear that states with $q>1.5$, produced via a moderate number of measurements, are indeed more advantageous in quantum sensing than a separable state.

\section{Conclusion and outlook}
We studied the relation between $\delta\omega$ and $q$ and derived an inequality $\delta\omega \leq O(N^{1-q})$. Any state with $q>1.5$ is advantageous in quantum metrology, hence we call states with $q>1.5$ SQL-beating catlike states.
We proposed a method to create SQL-beating catlike states \hk{through} repetitive measurements.
A thermal equilibrium state of the spin ensemble was coupled with an ancillary qubit, and we repeatedly measured the ancillary qubits. \hk{This} sequential measurement of the ancillary qubit provides information about the total magnetization of the spins, \hk{leading} the spin ensemble \hk{to} gradually approach %\ymdel{to} 
the SQL-beating catlike states.
\ym{
\hk{Notably, no} dynamical control over the spin ensembles \hk{is required} during the creation of
\hk{these} states.}
\hk{Analytically, we demonstrated} that the final state \hk{is} likely \hk{to} become
\ym{
\hk{a} SQL-beating catlike state} when the initial state is pure, i.e., \hk{at} zero temperature.
\hk{For} a mixed state at finite temperature as %\ymdel{an}
\hk{the} initial state,
\ym{we numerically showed that 
SQL-beating catlike states can be created via
repetitive measurements. }
We discussed the feasibility of our proposal concerning a hybrid system of an electron spin ensemble and superconducting flux qubit, strengthening the connection between quantum metrology and quantum computing.
These results pave the way for the realization of 
entanglement-enhanced metrology.

In the future work, it would be interesting to speed up the generation of SQL-beating catlike states, since cat states are fragile in general.
Recently, utilization of shortcuts to adiabaticity (STA) has been studied for quick generation of metrologically useful states, i.e., NOON states and spin squeezed states \cite{dengis2025multimode,odelli2023spin,odelli2024twist}.
Since our scheme includes measurements, it is nontrivial how we can speed up the generation. Still, since the STA formalism is widely studied and it is even extended to open quantum systems \cite{wu2021shortcuts}, it would be interesting to pursue the application of this formalism to our scheme.

\begin{acknowledgments}
M.T. was supported by the Japan Society for the Promotion of Science through a JSPS fellowship (JSPS KAKENHI Grant No. 22KJ3175).
This work was supported by JST Moonshot  (Grant Number JPMJMS226C). Y Matsuzaki was supported by JSPS KAKENHI (Grant Number 23H04390).
This work was supported by CREST (JPMJCR23I5, 20H05661), JST.
This work is supported by JST Moonshot (Grant Number JPMJMS226C) and Presto JST (JPMJPR245B).
R.H. was supported by JST ERATO Grant Number JPMJER2302, Japan, and JSPS KAKENHI Grant No. JP24K16982.
\as{A.S. was
supported by Japan Society for the Promotion of Science
KAKENHI Grant No. 23K22413 and by the RIKEN TRIP initiative.}
\end{acknowledgments}

\section*{Data Availability}
The data that support the findings of this article are openly 
available in Ref.~\footnote{10.5281/zenodo.18168860}.

\appendix
\section{Derivation of the state after the $(m+1)$th measurement}\label{stateaftermeas}
Let us investigate $\hat\rho_{\rm P}(m+1)$, the state after the $(m+1)$th measurement, from $\hat\rho_{\rm P}(m)$.
The measurement protocol is as follows.
First, we initialize the FQ so that the initial state is $\hat\rho_{\rm P}(m)\otimes \ket{+}\bra{+}$.
Then we let the total state evolve with the Hamiltonian $\hat H_{\mathrm R}=\frac{g}{2}\hat S_x\otimes \hat \Sigma_3$.
Finally, we read out the FQ on the $\hat\Sigma_2$ basis
($\hat \Sigma_2\ket{\pm_y}=\ket{\pm_y}$).
If we obtain $+1$ as an outcome of the measurement,
then $\hat\rho_{\rm P}(m+1)=\hat\rho_{\rm P}(m+1)_+$, which is written as
\begin{align}
&\hat\rho_{\rm P}(m+1)_+ \nonumber\\
&=
\frac{\braket{+_y|e^{-i\frac{g}{2}\hat S_x\otimes \hat \Sigma_3 t}\hat\rho_{\rm P}(m)\otimes |+}\bra{+}e^{i\frac{g}{2}\hat S_x\otimes \hat \Sigma_3 t}\ket{+_y}} 
{\mathrm{Prob}[\Sigma_2=+1]}
\\
&=\frac{\hat W_+\hat\rho_{\rm P}(m)\hat W_+^\dag}{\mathrm{Prob}[\Sigma_2=+1]},
\end{align}
where 
\begin{align}
\hat W_+&=\braket{+_y|e^{-i\frac{g}{2}\hat S_x\otimes \hat \Sigma_3 t}|+}\\
&=\frac{e^{-i\frac{g}{2}t\hat S_x}-ie^{i\frac{g}{2}t\hat S_x}}{2}\\
&=\frac{1-i}{\sqrt{2}}\sin\left(\frac{\pi}{4}+\frac{gt}{2}\hat S_x\right)
\end{align}
and
\begin{align}
&\mathrm{Prob}[\Sigma_2=+1]\nonumber\\
&=\mathrm{Tr}\left(
\braket{+_y|e^{-i\frac{g}{2}\hat S_x\otimes \hat \Sigma_3 t}\hat\rho_{\rm P}(m)\otimes |+}\bra{+}
e^{i\frac{g}{2}\hat S_x\otimes \hat \Sigma_3 t}\ket{+_y}
\right).
\end{align}

Similarly, if we obtain $-1$ as an outcome of the measurement,
then $\hat\rho_{\rm P}(m+1)=\hat\rho_{\rm P}(m+1)_-$, which is written as
\begin{align}
&\hat\rho_{\rm P}(m+1)_-\nonumber\\
&=
\frac{\braket{-_y|e^{-i\frac{g}{2}\hat S_x\otimes \hat \Sigma_3 t}\hat\rho_{\rm P}(m)\otimes |+}\bra{+}
e^{i\frac{g}{2}\hat S_x\otimes \hat \Sigma_3 t}\ket{-_y}}
{\mathrm{Prob}[\Sigma_2=-1]}
\\
&=\frac{\hat W_-\hat\rho_{\rm P}(m)\hat W_-^\dag}{\mathrm{Prob}[\Sigma_2=-1]},
\end{align}
where 
\begin{align}
\hat W_-&=\braket{-_y|e^{-i\frac{g}{2}\hat S_x\otimes \hat \Sigma_3 t}|+}\\
&=\frac{e^{-i\frac{g}{2}t\hat S_x}+ie^{i\frac{g}{2}t\hat S_x}}{2}\\
&=\frac{1+i}{\sqrt{2}}\sin\left(\frac{\pi}{4}-\frac{gt}{2}\hat S_x\right)
\end{align}
and
\begin{align}
&\mathrm{Prob}[\Sigma_2=-1]\nonumber\\
&=\mathrm{Tr}\left(
\braket{-_y|e^{-i\frac{g}{2}\hat S_x\otimes \hat \Sigma_3 t}\hat\rho_{\rm P}(m)\otimes |+}\bra{+}
e^{i\frac{g}{2}\hat S_x\otimes \hat \Sigma_3 t}\ket{-_y}
\right).
\end{align}

\section{Derivation of $\ket{\phi_m}$ }\label{appB}
Here let us discuss the state we get after applying $\hat W_+^k\hat W_-^{m-k}$, where  $0\leq k\leq m$ is the number of $\hat W_+$ applied in $m$ measurements. 
There is a theorem that when an %Kraus 
operator $\hat K$ is applied infinite times, the state converges to the eigenstate of $\hat K$ with the  eigenvalue whose modulus is the largest, provided it is nondegenerate.
Furthermore, 
in our case, for $k=\alpha m$ where $0<\alpha <1$, we obtain $(\hat W_+)^k(\hat W_-)^{m-k}=((\hat W_+)^{\alpha }(\hat W_-)^{(1-\alpha)})^m$. 
By taking $\hat K:= (\hat W_+)^{\alpha }(\hat W_-)^{(1-\alpha)}$, we can adopt this theorem.

We want to find the largest (in magnitude) eigenvalue of  
\begin{align}
&\hat K
\nonumber\\
&=\left(\frac{1-i}{\sqrt{2}}\sin(\frac{\pi}{4}+\frac{gt}{2}\hat S_x)\right)^\alpha
\left(\frac{1+i}{\sqrt{2}}\sin(\frac{\pi}{4}-\frac{gt}{2}\hat S_x)\right)^{1-\alpha}
\end{align}
 for each $\alpha$.
 When 
$\hat K$
is applied to some initial state $\ket{\psi_0}=\sum_{l}c_l\ket{S_x=l}$, then
\begin{align}
&\hat K
\ket{\psi_0}
=
\left(\frac{1-i}{\sqrt{2}}\right)^\alpha
\left(\frac{1+i}{\sqrt{2}}\right)^{1-\alpha}\nonumber\\
&\times
\sin^\alpha\left(\frac{\pi}{4}+\frac{gt}{2}\hat S_x\right)
\sin^{1-\alpha}\left(\frac{\pi}{4}-\frac{gt}{2}\hat S_x\right)
\sum_{l}c_l\ket{S_x=l}\\
&=
\left(\frac{1-i}{\sqrt{2}}\right)^\alpha
\left(\frac{1+i}{\sqrt{2}}\right)^{1-\alpha}\nonumber\\
&\times
\sum_{l}c_l\sin^\alpha\left(\frac{\pi}{4}+\frac{gt}{2}l\right)
\sin^{1-\alpha}\left(\frac{\pi}{4}-\frac{gt}{2}l\right)\ket{S_x=l}.
\end{align}
 Let us define $f(x)$ as
\begin{align}
f(x):=\sin^\alpha\left(\frac{\pi}{4}+\frac{x}{2}\right)
\sin^{1-\alpha}\left(\frac{\pi}{4}-\frac{x}{2}\right).
\end{align}
Finding the extremum of $f(x)$ corresponds to finding the largest eigenvalue of
%$\hat W_+^k\hat W_-^{m-k}$ 
$\hat K$
because $\hat W_{\pm}$ are diagonal in the $\hat S_x$ basis.

To find the extremum, we take the derivative
\begin{align}
f'(x)
=\frac{2\alpha-1-\sin x
}{4\sin^{1-\alpha}\left(\frac{\pi}{4}+\frac{x}{2}\right)\sin^{\alpha}\left(\frac{\pi}{4}-\frac{x}{2}\right)}.
\end{align}
We can see that $f(x)$ takes the extremum at $x$ that satisfies $g(x)=0$,
where
\begin{align}
g(x):=2\alpha-1-\sin x.
\end{align}

It is obvious that $g(x)$ takes zero at
\begin{align}
\sin x = 2\alpha-1.%\frac{2k}{m}-1.
\end{align}
For simplicity, let us first consider the region 
\begin{align}
-\frac{\pi}{2} < x< \frac{\pi}{2}.
\end{align}
Since $0<\alpha< 1$,
 $f(x)$ takes the maximum at $x=\arcsin(2\alpha-1)$, which is greater than zero and smaller than $\pi/2$. 
When $\alpha= 1$ ($-1$), $f(x)$ takes the maximum (minimum) $1$ ($-1$) at $x=\pi/2$ ($-\pi/2$).
These are illustrated as in Table \ref{zougenhyou}.

\begin{table}[h]
    \centering
    \caption{Derivative test chart of $f(x)$, which increases until $x=\arcsin(2\alpha-1)$ and then decreases.}
    \label{zougenhyou}
    \begin{tabular}{cccc}
        \hline \hline
        $x$ & $-\pi/2 $ & $\arcsin(2\alpha-1)$ & $\pi/2 $ \\ \hline
        $f'(x)$ & $+$  & $0$ & $-$ \\ 
$f(x)$ & $\nearrow$ &  &$\searrow$ \\ 
 \hline\hline
    \end{tabular}
\end{table}

Therefore, under the assumption $-\pi/2 < x < \pi/2$, the function $f(x)$ takes the maximum at $x=\arcsin(2\alpha-1)$,
leading to the conclusion that after applying $\hat K$ for $m \gg 1$ times, the final state  $\ket{\phi_m}$ approaches (\ref{noninteger}).

\ym{Let us also consider the case of}
%How about in the region 
$x\geq \pi/2$ or $x\leq-\pi/2$.
In general, $g(x)$ takes zero at
\begin{align}
x=\arcsin(2\alpha-1)+2n\pi,\\
x=\pi-\arcsin(2\alpha-1)+2n\pi,
\end{align}
where $n\in \mathbb{Z}$.
Actually, we can prove that 
\begin{align}
|f(x)|
&=\left(1-\alpha\right)^{\frac{1-\alpha}{2}}
\left(\alpha\right)^{\frac{\alpha}{2}}\\
&=\sqrt{(1-\alpha)^{1-\alpha}\alpha^\alpha}
\end{align}
for any $x$ that satisfies $g(x)=0$, i.e., $\sin x=2k/m-1$.
With the extrema with the equal value of $|f(x)|$, this implies that there are multiple candidates for the final state.
In such a case, we should consider the weight of the eigenstates that are candidates for the final state, in the initial state.
in other words, the final state should be the superposition of the candidates whose  weight is determined by the initial state.

\section{Derivation of $p(k)$ }\label{appProb}
\ym{Here, we explain the derivation of $p(k)$,} the probability of obtaining the trajectory with $\hat W_+$ applied $k$ times.
\begin{align}
&\left(\hat W_+^\dag\hat W_+\right)^k
\left(\hat W_-^\dag\hat W_-\right)^{m-k}\nonumber\\
&=
\left(\sin^2(\frac{\pi}{4}+\frac{gt}{2}\hat S_x)\right)^k
\left(\sin^2(\frac{\pi}{4}-\frac{gt}{2}\hat S_x)\right)^{m-k}\\
&=
\left(\frac{1+\sin(gt\hat S_x)}{2}\right)^k
\left(\frac{1-\sin(gt\hat S_x)}{2}\right)^{m-k}.
\end{align}
We use  $\ket{S_x=2r-N}$ (where $r=0, 1,..., N$ is the number of up spins) in (\ref{dicke}) as
\begin{align}
&\ket{S_x=2r-N}
=\ket{D_N^{(\theta)}}\nonumber\\
&=\sqrt{\binom{N}{\frac{N+\theta}{2}}^{-1}}\sum_{\sigma\in \mathcal{S}_N}
\mathcal{P}_\sigma\left(\ket{+}^{\otimes (N+\theta)/2}\ket{-}^{\otimes (N-\theta)/2}\right)
\end{align}
with $\theta=2r-N$.
Importantly,
\begin{align}
\ket{\uparrow}^{\otimes N}
&=
\left(\frac{\ket{+}+\ket{-}}{\sqrt{2}}\right)^{\otimes N}\nonumber\\
&=
\frac{1}{\sqrt{2^N}}\sum_{r=0}^N\ket{S_x=2r-N}\sqrt{\binom{N}{r}}.
\end{align}
Hence we obtain
\begin{align}
&p(k)
%\\
%&=
=
\binom{m}{k}
\bra{\uparrow}^{\otimes N}
\left(\hat W_+^\dag\hat W_+\right)^k
\left(\hat W_-^\dag\hat W_-\right)^{m-k}
\ket{\uparrow}^{\otimes N}
\\
&=
%\binom{m}{k}
%\bra{\uparrow}^{\otimes N}
%\left(\frac{1+\sin(gt\hat S_x)}{2}\right)^k
%\left(\frac{1-\sin(gt\hat S_x)}{2}\right)^{m-k}
%\ket{\uparrow}^{\otimes N}
\binom{m}{k}
\frac{1}{2^N}
\sum_{r=0}^N\bra{ S_x=2r-N}\sqrt{\binom{N}{r}}\nonumber\\
&
\times
\left(\frac{1+\sin(gt\hat S_x)}{2}\right)^k
\left(\frac{1-\sin(gt\hat S_x)}{2}\right)^{m-k}\nonumber\\
&
\times
\sum_{r'=0}^N\ket{S_x=2r'-N}\sqrt{\binom{N}{r'}}\\
&=
\binom{m}{k}
\frac{1}{2^N}
\sum_{r=0}^N
\binom{N}{r}\nonumber\\
&\times
\left(\frac{1+\sin(gt(2r-N))}{2}\right)^k
\left(\frac{1-\sin(gt(2r-N))}{2}\right)^{m-k}.
\end{align}

\section{Probability of obtaining a generalized cat state}\label{appProb1}
We consider the case where   $m$ is even.
Here we discuss the probablity of obtaining a generalized cat state after $m$ measurements with $\ket{\uparrow}^{\otimes N}$ as an initial state.
We take the sum of $p(k)$ over $k$, which allows the emergence of generalized cat states, and prove that it converges to $1$ in the limit $N\rightarrow \infty$ (note that we take $N\rightarrow\infty$ after $m\rightarrow\infty$). 
The probability $p(k)$ can be regarded as a sum of binomial distributions with weight $\binom{N}{r}/2^N$,
\begin{align}
p(k)
&=\frac{1}{2^N}
\sum_{r=0}^N
\binom{N}{r}
B(gt,r,k),\\
B(gt,r)&:=\binom{m}{k}s(gt,r)^k(1-s(gt,r))^{m-k},\\
s(gt,r)&:=\frac{1+\sin(gt(2r-N))}{2}.\label{sgtr}
\end{align}
%The assumption $gtN<\pi/2$ makes 
If we assume $gtN<\pi/2$, we find 
$s(gt,r)>0$ and $1-s(gt,r)>0$.
Since we are considering large $m$, the binomial distribution $B(gt,r)$ can be approximated as the normal distribution, i.e.,
\begin{align}
&B(gt,r,k)
\simeq \frac{1}{\sqrt{2\pi ms(gt,r)(1-s(gt,r))}}\nonumber\\
&\times\exp\left(-\frac{(k-ms(gt,r))^2}{2ms(gt,r)(1-s(gt,r))}\right).\\
&\therefore
p(k)\simeq \frac{1}{2^N}
\sum_{r=0}^N
\binom{N}{r}
\frac{1}{\sqrt{2\pi ms(gt,r)(1-s(gt,r))}}\nonumber\\
&\times\exp\left(-\frac{(k-ms(gt,r))^2}{2ms(gt,r)(1-s(gt,r))}\right)
\end{align}
Note that $s(gt,r)$ is a monotonically increasing function of $r$.

Consider the sum of $p(k)$ from $k=m\alpha_1$ to $k=m\alpha_2$.
When $m\gg1 $, we can replace the sum with integral, i.e.,
\begin{align}
&\sum_{k=m\alpha_1}^{m\alpha_2}p(k)\nonumber\\
&= \sum_{k=m\alpha_1}^{m\alpha_2}\frac{1}{2^N}\sum_{r=0}^N\binom{N}{r}B(gt,r,k)\\
&\simeq \frac{1}{2^N}\sum_{r=0}^N\binom{N}{r}\sqrt{\frac{m}{2\pi s(gt,r)(1-s(gt,r))}}
\nonumber\\
&\times 
\int_{\alpha_1}^{\alpha_2}d\alpha \exp\left(-\frac{(\alpha-s(gt,r))^2}{2s(gt,r)(1-s(gt,r))}m\right)\\
&\rightarrow
\frac{1}{2^N}\sum_{r=0}^N\binom{N}{r}
\int_{\alpha_1}^{\alpha_2}d\alpha \delta(\alpha-s(gt,r))
\quad (m\rightarrow \infty)
\end{align}
Here we substitute $\gamma=\sqrt{\frac{s(gt,r)(1-s(gt,r))}{m}}$ into the following known formula:
\begin{align}
\lim_{\gamma\rightarrow 0}\frac{\exp\left(-\frac{x^2}{2\gamma^2}\right)}{\sqrt{2\pi}\gamma}=\delta(x)
\end{align}
Let us take $\alpha_1=s(gt,r_1)$ and $\alpha_2=s(gt,r_2)$. Then the integral is further simplified as follows:
\begin{align}
&\sum_{k=ms(gt,r_1)}^{ms(gt,r_2)}p(k)%\nonumber\\
\simeq \frac{1}{2^N}\sum_{r=r_1}^{r_2}\binom{N}{r}.
\end{align}

Now, to evaluate the lower bound of the probability of the emergence of the generalized cat states, 
we can take $r_1=Nx_1$ and $r_2=Nx_2$ with $0<x_1<1/2<x_2<1$ in the above equation.
To see that this choice is appropriate, we  substitute
$k=ms(gt,\tilde{r})$ to Eq.~(\ref{noninteger}), finding
\begin{align}
\ket{\phi_m}=\ket{S_x=2\tilde{r}-N}.
\end{align}
According to Eq.~(\ref{nekojouken}), this means that the final state $\ket{\phi_m}$ is indeed a generalized cat state for any  $\tilde{r}=Nx$ with $0<x<1$.

Then, we finally have
\begin{align}
&\sum_{k=ms(gt,r_1)}^{ms(gt,r_2)} p(k)\\
&=\frac{1}{2^N}\sum_{r=Nx_1}^{Nx_2}\binom{N}{r}\\
&\simeq \sum_{r=Nx_1}^{Nx_2}\sqrt{\frac{2}{\pi N}}\exp\left(-\frac{2(r-N/2)^2}{N}\right)\\
&\rightarrow 
\int_{x_1}^{x_2}dx \delta(x-1/2)
\quad (N\rightarrow \infty).
\end{align}
Here, we approximated the binomial distribution in the same manner as we did with $B(gt,r,k)$, assuming $N \gg 1$.
We can immediately see that this approaches $1$ in the limit $N\rightarrow \infty$.
Therefore, the probability of obtaining a generalized cat state after infinitely many measurements is $1$ with $N\rightarrow \infty$.

\bibliography{refnvandfq}% Produces the bibliography via BibTeX.

\end{document}